\documentclass[preprint]{elsarticle}
\usepackage{amssymb}
\usepackage[final]{epsfig}
\usepackage{color}
\usepackage{graphicx}
\usepackage{lineno}
\usepackage{bm}
\newcommand{\beq}{\begin{equation}}
\newcommand{\eeq}{\end{equation}}
\newcommand{\bdi}{\begin{displaymath}}
\newcommand{\edi}{\end{displaymath}}
\newcommand{\no}{\nonumber}
\newcommand{\bea}{\begin{eqnarray}}
\newcommand{\eea}{\end{eqnarray}}
\newcommand{\vep}{\varepsilon}
\newcommand{\ov}{\overline}
\newcommand{\grad}{{\bf \nabla}}
\newcommand{\vr}{{\bf r}}
\newcommand{\vv}{{\bf v}}

\newcommand{\vE}{{\bf E}}
\newcommand{\vK}{{\bf K}}
\newcommand{\vB}{{\bf B}}
\newcommand{\vD}{{\bf D}}
\newcommand{\vH}{{\bf H}}
\newcommand{\vF}{{\bf F}}
\newcommand{\vG}{{\bf G}}
\newcommand{\vA}{{\bf A}}
\newcommand{\vx}{{\bf x}}
\newcommand{\vJ}{{\bf J}}
\newcommand{\vbeta}{{\bm  \beta}}

\newcommand{\vb}{{\bf b}}
\newcommand{\msigma}{{\bf \hat \sigma}}
\newcommand{\mvep}{{\bf \hat \varepsilon}}
\newcommand{\mmu}{{\bf \hat \mu}}
\newcommand{\ep}{\varepsilon}

\newcommand{\tret}{t_{\mathrm{ret}}}

\newcommand{\text}{}

\newcommand{\de}{\partial}
\newcommand{\om}{\omega}

\begin{document}

\begin{frontmatter}

\title{Signals induced on electrodes by moving charges, a general theorem for Maxwell's equations based on Lorentz-reciprocity}

\author[CERN]{W. Riegler}
\author[Oxford]{P. Windischhofer}

\address[CERN]{CERN}
\address[Oxford]{University of Oxford}

\begin{abstract}
  
We discuss a signal theorem for charged particle detectors where the finite propagation time of the electromagnetic waves produced by a moving charge cannot be neglected. While the original Ramo-Shockley theorem and related extensions are all based on electrostatic or quasi-electrostatic approximations, the theorem presented in this report is based on the full extent of Maxwell's equations and does account for all electrodynamic effects. It is therefore applicable to all devices that detect fields and radiation from charged particles. 

\end{abstract}

\end{frontmatter}

\section{Introduction}

Ever since the publication of Shockley \cite{shockley} and Ramo \cite{ramo} discussing the theorems for induction of currents on grounded electrodes by moving charges, there have been efforts to extend these relations to more general situations: geometries including space-charge \cite{gatti}, signals on electrodes connected with impedance elements \cite{radeka}, formulation of equivalent circuits \cite{blum}, inclusion of permittivity and non-linear materials \cite{hamel1, hamel2} as well as geometries that contain material of finite resistivity \cite{werner1, werner2, werner3}. All of these extensions are based on static or quasi-static approximations of Maxwell's equations together with Green's reciprocity theorem relating electro-static potentials and charge distributions. They apply to situations where the finite propagation velocity of electromagnetic waves, and indeed all radiation phenomena, can be neglected. \\ 
However, e.g. for detectors with long readout electrodes, the finite propagation velocity of signals plays a significant role. This situation is typically treated by calculating the induced signal using the Ramo-Shockley theorem and then placing the signal as an ideal current source on the transmission line at the place of the charge movement \cite{werner4}. While 'intuitively' this is the correct solution, it is interesting to investigate whether more general reciprocity theorems that are valid for the full extent of Maxwell's equations can be directly applied to this situation. \\
We can also view antennas that detect the radiation from moving charges to be 'electrodes' on which the signal is induced. A generalized theorem might also be applied to this situation. It turns out that the Lorentz reciprocity theorem \cite{lorentz} can indeed be used to derive such a very general signal theorem, where a weighting field $\vE_w(\vx, t)$ of the electrode in question is used to calculate the signal. 
\\ 
This report is structured as follows. We first outline the Lorentz reciprocity theorem and two immediate consequences, namely the network reciprocity theorem and the antenna reciprocity theorem. Then we derive the generalized signal theorem and apply it to four specific examples, namely transmission lines, synchrotron radiation from gyrating electrons, signals in beam current transformers and signals arising form the Askaryan effect. Two appendices show the explicit equivalence of the direct signal calculation from the Lienard-Wiechert potentials and the calculation using weighting fields, for the cases of infinitesimal electric and magnetic dipole antennas.

%%%%%%%%%%%%%%%%%%%%%%%%%%%%%%%%%%%%%%%%%%%%%%%%%%%%%%%%%%%
%%%%%%%%%%%%%%%%%%%%%%%%%%%%%%%%%%%%%%%%%%%%%%%%%%%%%%%%%%%

\section{Lorentz reciprocity theorem} 

\begin{figure}[ht]
 \begin{center}
 a)
  \includegraphics[width=6cm]{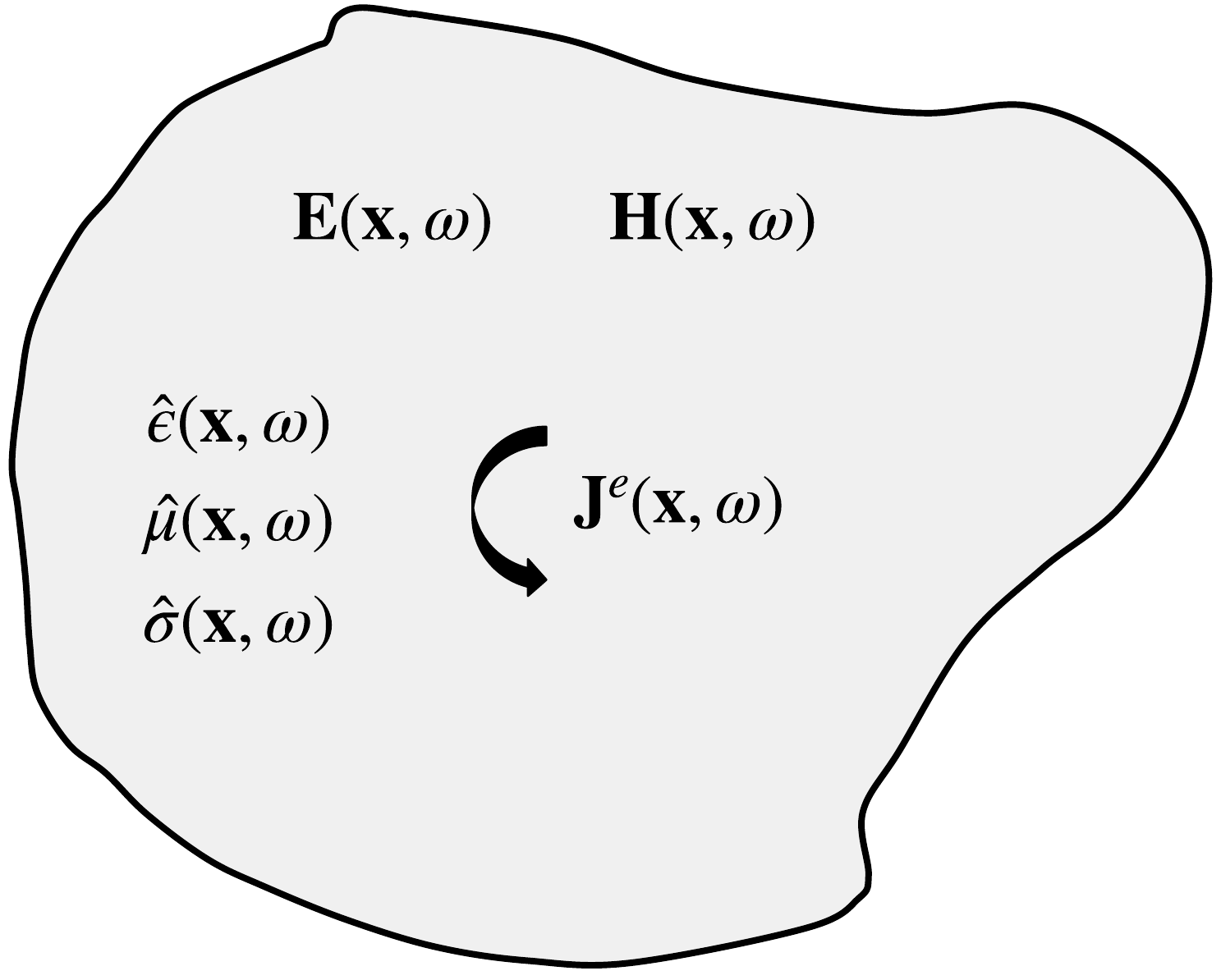}
 b)
    \includegraphics[width=6cm]{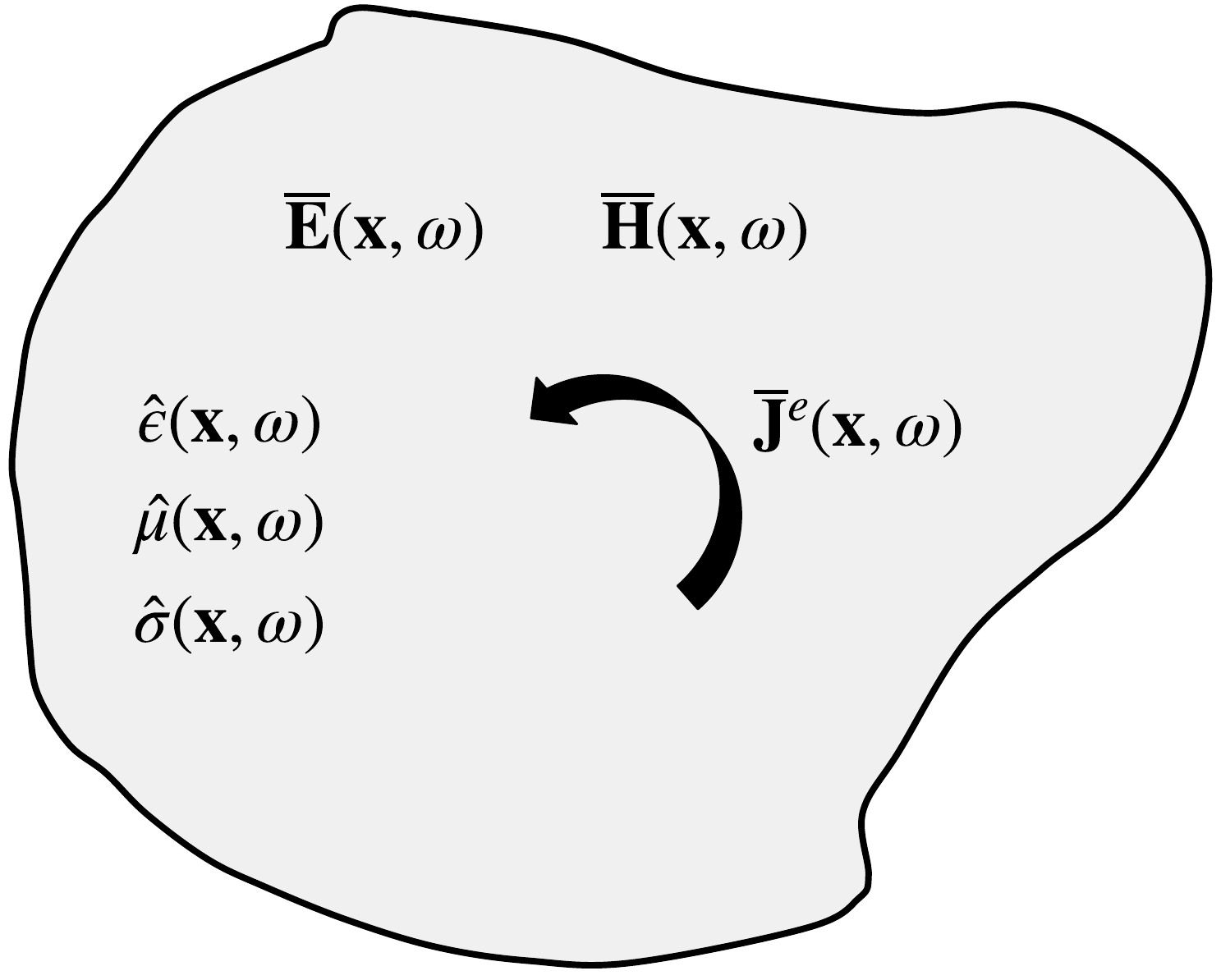}
  \caption{Two different current densities in the same geometry with the same material properties.}
  \label{theorem0}
  \end{center}
\end{figure}

We assume the most general form of Maxwell's equations for a linear anisotropic material of position- and frequency-dependent permittivity matrix $\mvep(\vx, \om)$, permeability matrix $\mmu (\vx, \om)$ and conductivity matrix $\msigma (\vx, \om)$. These $3\times 3$ matrices relate the vector fields
\beq
       \vD = \hat \vep \vE \qquad  \vB = \hat \mu \vH \qquad \vJ = \hat \sigma \vE
\eeq
The source of the fields is an externally impressed current density $\vJ^e(\vx, \om)$. In the Fourier domain, Maxwell's equations then read as
\beq
     \grad \cdot \hat \vep \vE = \rho \qquad  \grad \cdot \hat \mu \vH = 0
\eeq
\beq
    \grad\times \vE = -i \om \hat \mu \vH  \qquad \grad \times \vH = \vJ^e+\msigma \vE + i\om  \hat \vep \vE
\eeq
Let us now look at the situation where two different externally impressed current densities $\vJ^e$ and $\ov \vJ^e$ are placed on the same material distribution, as shown in Fig.~\ref{theorem0}.
The current density $\vJ^e$ will cause fields $\vE$ and $\vH$ and the current density $\ov \vJ^e$ will cause fields $\ov \vE$ and $\ov \vH$. 
We can relate these quantities using the following expression, which holds for two general vector fields $\vF$ and $\vG$:
\beq
     \grad\cdot (\vF \times \vG ) = \vG\cdot(\grad \times \vF) - \vF\cdot(\grad \times \vG)
\eeq
We can therefore write
\bea
    \grad\cdot (\vE \times \ov \vH) &  = &\ov \vH(\grad \times \vE) - \vE(\grad \times \ov \vH)  \\
       & = & - \vE \ov \vJ^e -i \om \ov \vH \mmu \vH - \vE(\msigma+i\om\mvep)  \ov \vE \\
           \grad\cdot (\ov \vE \times \vH) &  = & \vH(\grad \times \ov \vE) - \ov \vE(\grad \times \vH)  \\
       & = & - \ov \vE \vJ^e -i \om \vH \mmu \ov \vH - \ov \vE(\msigma+i\om\mvep)  \vE
\eea
Also, for any two vectors $\vF$ and $\vG$ it holds that
\beq
   {\bf F \hat m \bf G} =   {\bf G \hat m \bf F} \qquad \mbox{if} \qquad {\bf \hat m} = {\bf \hat m}^T
\eeq
By subtracting the two expressions from above and assuming $\mvep, \mmu$ and $\msigma$ to be symmetric we get
\beq
     \grad (\vE \times \ov \vH  - \ov \vE \times \vH)  =  \ov \vE \vJ^e - \vE \ov \vJ^e
\eeq
Integrating over a volume $V$ enclosed by surface $A$ and applying Gauss' theorem we have
\beq
     \oint_A  (\vE \times \ov \vH  - \ov \vE \times \vH) d \vA = \int_V ( \ov \vE \vJ^e - \vE  \ov \vJ^e) dV
\eeq
If the sources are compact i.e. all sources are contained in a finite region of space, the left hand side evaluates to zero. This is true even in the presence of electromagnetic radiation propagating towards spatial infinity \cite{stumpf}. We thus get the Lorentz reciprocity theorem:
\beq \label{lorentz_reciprocity_theorem}
    \int_V \ov \vE(\vx, \om) \vJ^e (\vx, \om)dV = \int_V \vE (\vx, \om) \ov \vJ^e(\vx, \om) dV
\eeq
It is important to appreciate that $\vJ^e(\vx, \om)$ refers to the 'externally impressed' currents. In case the conductivity matrix $\hat \sigma(\vx)$ is different from zero there will be additional currents according to $\vJ = \hat \sigma \vE$. This does however not change the form of the theorem where only the external currents appear. The above relation holds only for $\vJ^e \neq 0$ and $\ov \vJ^e \neq 0$. If one of the two currents is equal to zero the relation simplifies to $0=0$ and has therefore no useful application.

%%%%%%%%%%%%%%%%%%%%%%%%%%%%%%%%%%%%%%%%%%%%%%%%%%%%%%%%%%%

\subsection{Network reciprocity}
\label{network_reciprocity}

\begin{figure}[ht]
 \begin{center}
 a)
  \includegraphics[width=6cm]{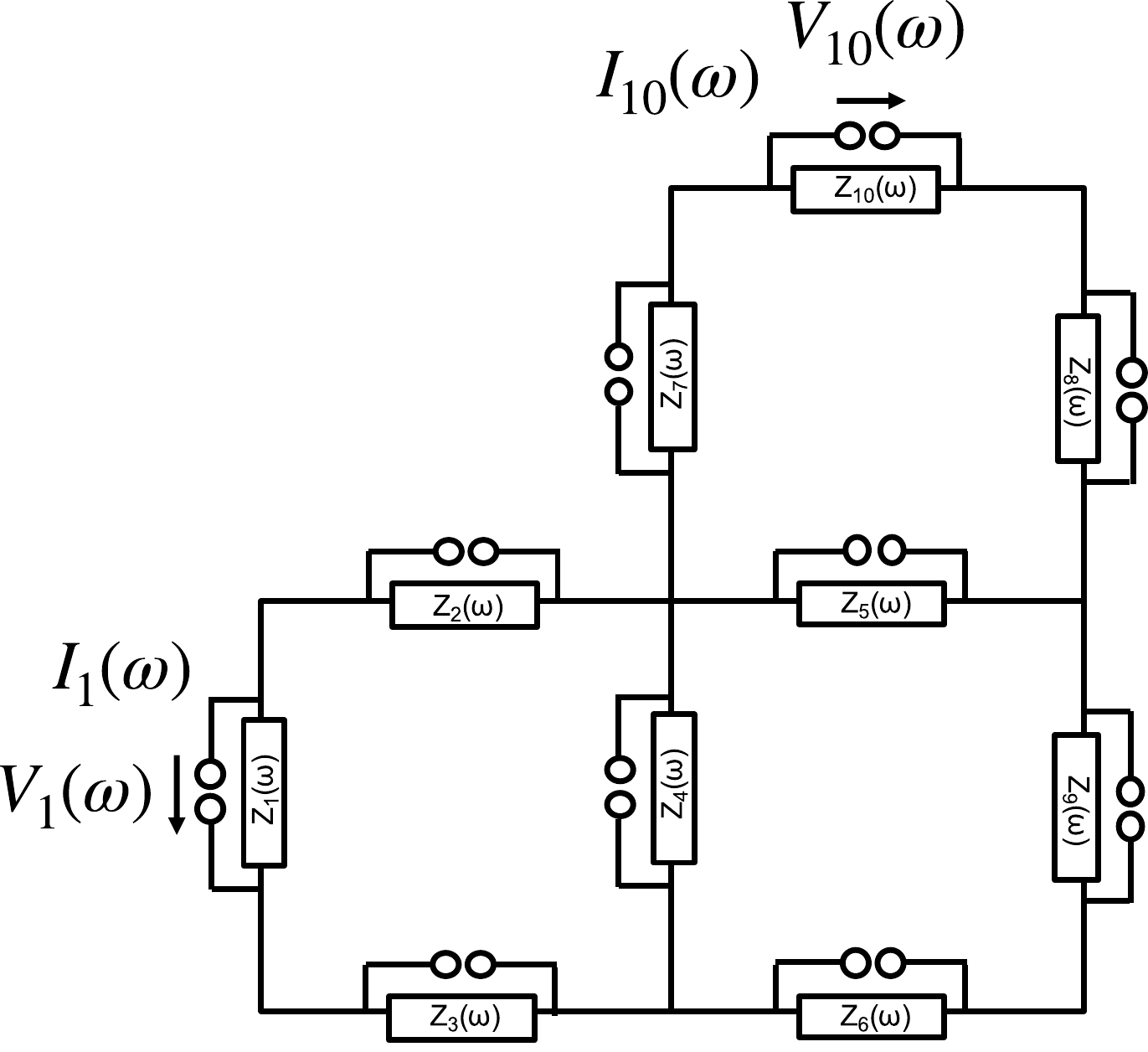}
 b)
    \includegraphics[width=6cm]{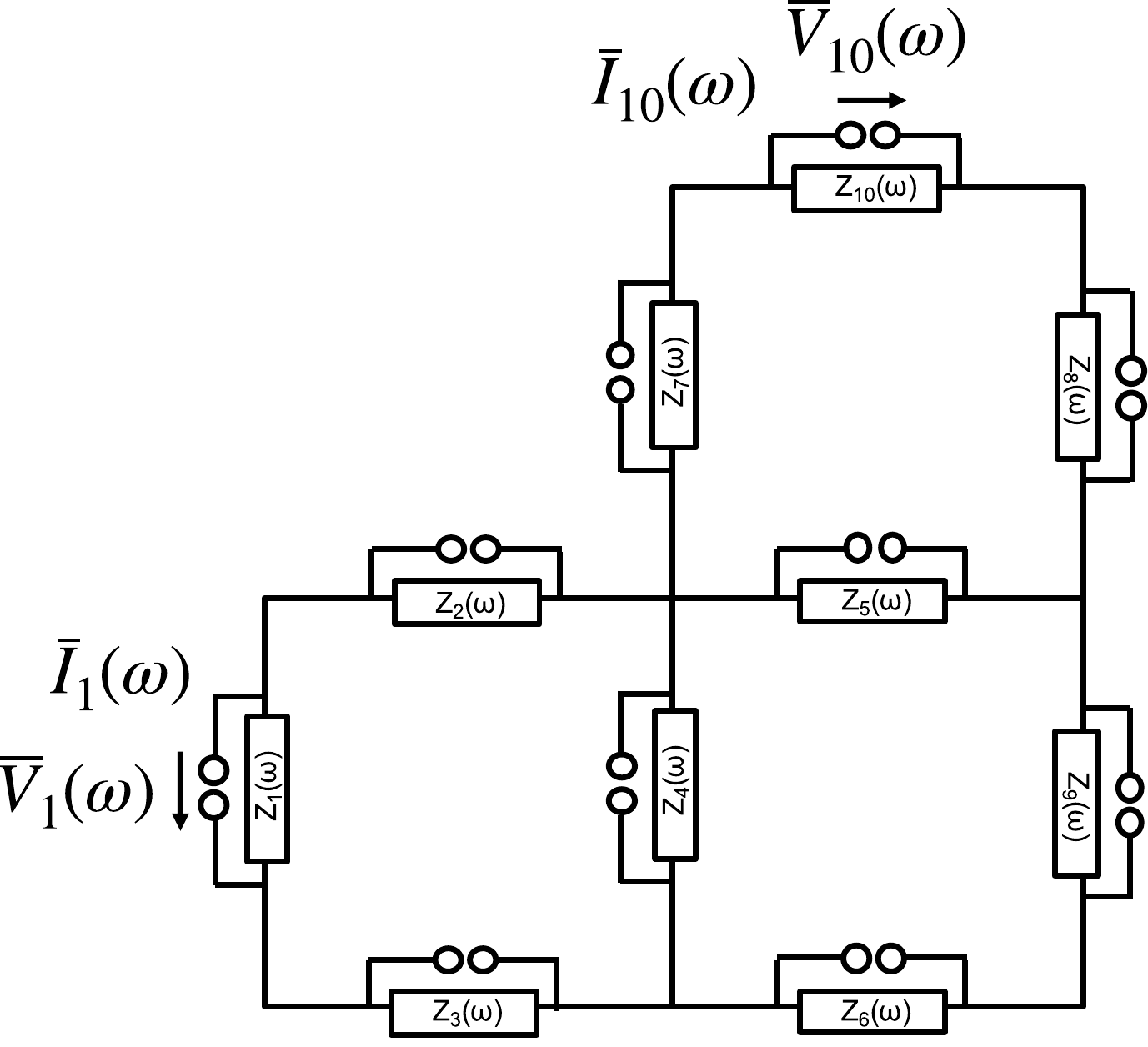}
  \caption{A network of impedance elements with localized current sources. a) The current sources $I_n(\om)$ produce voltages $V_n(\om)$ b) The current sources $\ov I_n(\om)$ produce voltages $\ov V_n(\om)$.  }
  \label{network_reciprocity}
  \end{center}
\end{figure}

The fact that the Lorentz reciprocity theorem holds for arbitrary $\hat \vep(\vx,\om), \hat \mu(\vx,\om), \hat \sigma(\vx,\om)$ allows us to assume that a general linear reactive network made from discrete elements and connected with wires is just a specific realization of these material distributions, as shown in Fig.~\ref{network_reciprocity}.  In Fig.~\ref{network_reciprocity}a) we connect external localized 'point-current' sources $I_n(\om)$ to the system. The resulting electric fields between the two open terminals are related to the measured voltages $V_n(\om)=\vE_n( \om)d{\bf s}$. 
In the second situation in Fig.~\ref{network_reciprocity}b we use a different set of currents $\ov I_n(\om)$ resulting in voltages $\ov V_n(\om)$. Using the reciprocity theorem with the current sources $\vJ_n(\vx, \om)=I_n(\om)\delta(\vx-\vx_n) d{\bf s}$ etc. and resulting voltages $V_n(\om) = \vE(\vx, \om) d{\bf s}$  we find
\beq
       \sum_{n=1}^{N=10} I_n(\om) \ov V_n(\om) =  \sum_{n=1}^{N=10} \ov I_n(\om) V_n(\om) 
\eeq
This expression is the counterpart of the electrostatic reciprocity theorem, where different sets of charges and voltages on metal electrodes are related by  $\sum_{n=1}^N Q_n \ov V_n=  \sum_{n=1}^N \ov Q_n V_n$. \\ \\
We now allow only a single current source to deliver a nonvanishing current $I_1(\om)$, while all other sources are switched off. Likewise, we place only a single current $\ov I_{10}(\om)$ and no currents across the other sources. We then get
\beq
       I_1(\om) \ov V_1(\om) = \ov I_{10}(\om) V_{10}(\om) 
       \quad \rightarrow \quad
       \frac{I_1(\om)}{V_{10}(\om)} =  \frac{\ov I_{10}(\om)}{\ov V_1(\om)} 
\eeq
This is the network reciprocity theorem: \\
{\it The voltage across an impedance element $Z_m(\om)$ due to a current $I(\om)$ on a different element $Z_n(\om)$ is equal to the voltage across  $Z_n(\om)$ for the same current  $I(\om)$ on $Z_m(\om)$.}

%%%%%%%%%%%%%%%%%%%%%%%%%%%%%%%%%%%%%%%%%%%%%%%%%%%%%%%%%%%

\subsection{Antenna reciprocity}

\begin{figure}[ht]
 \begin{center}
 a)
  \includegraphics[width=6cm]{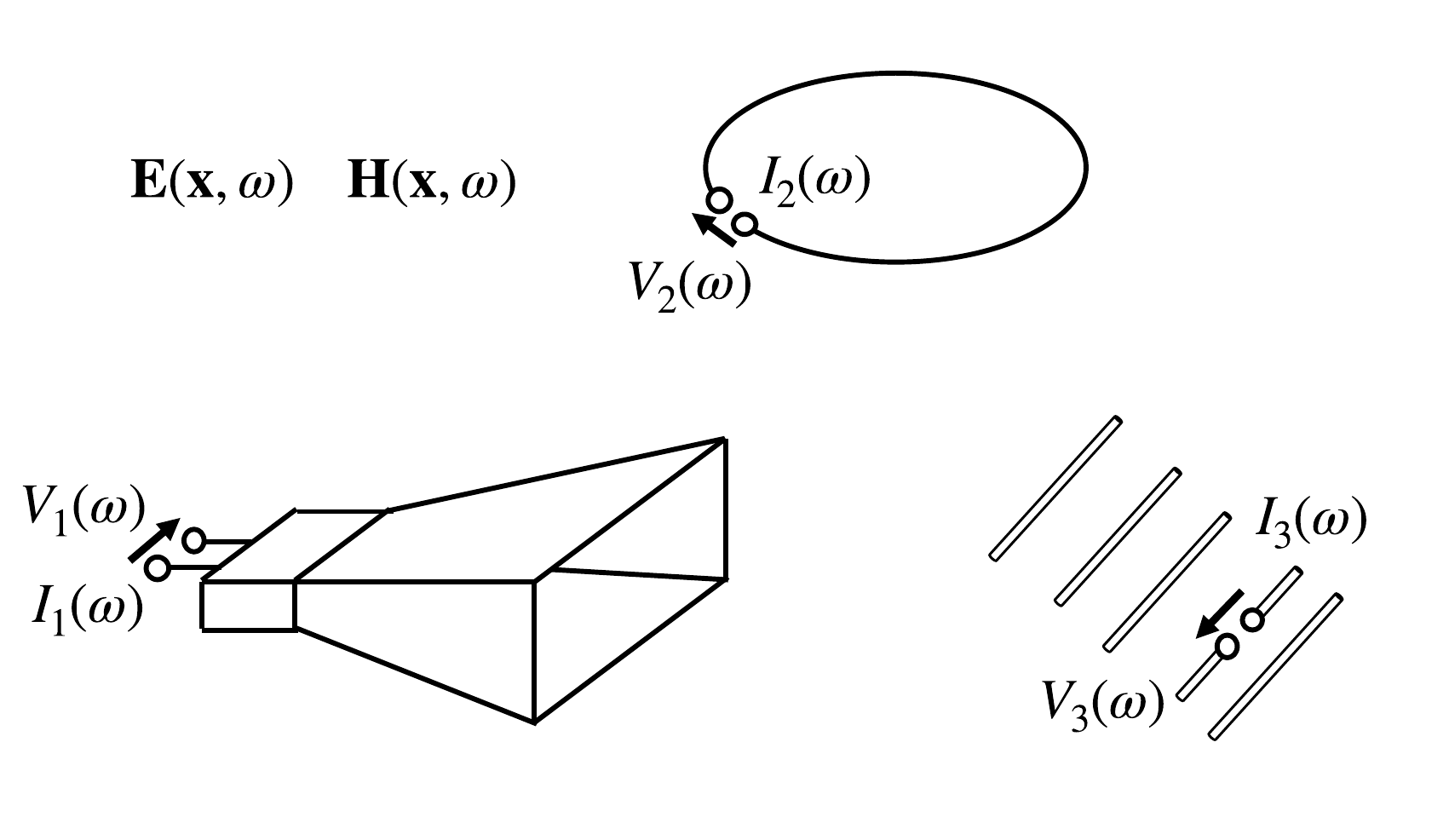}
 b)
    \includegraphics[width=6cm]{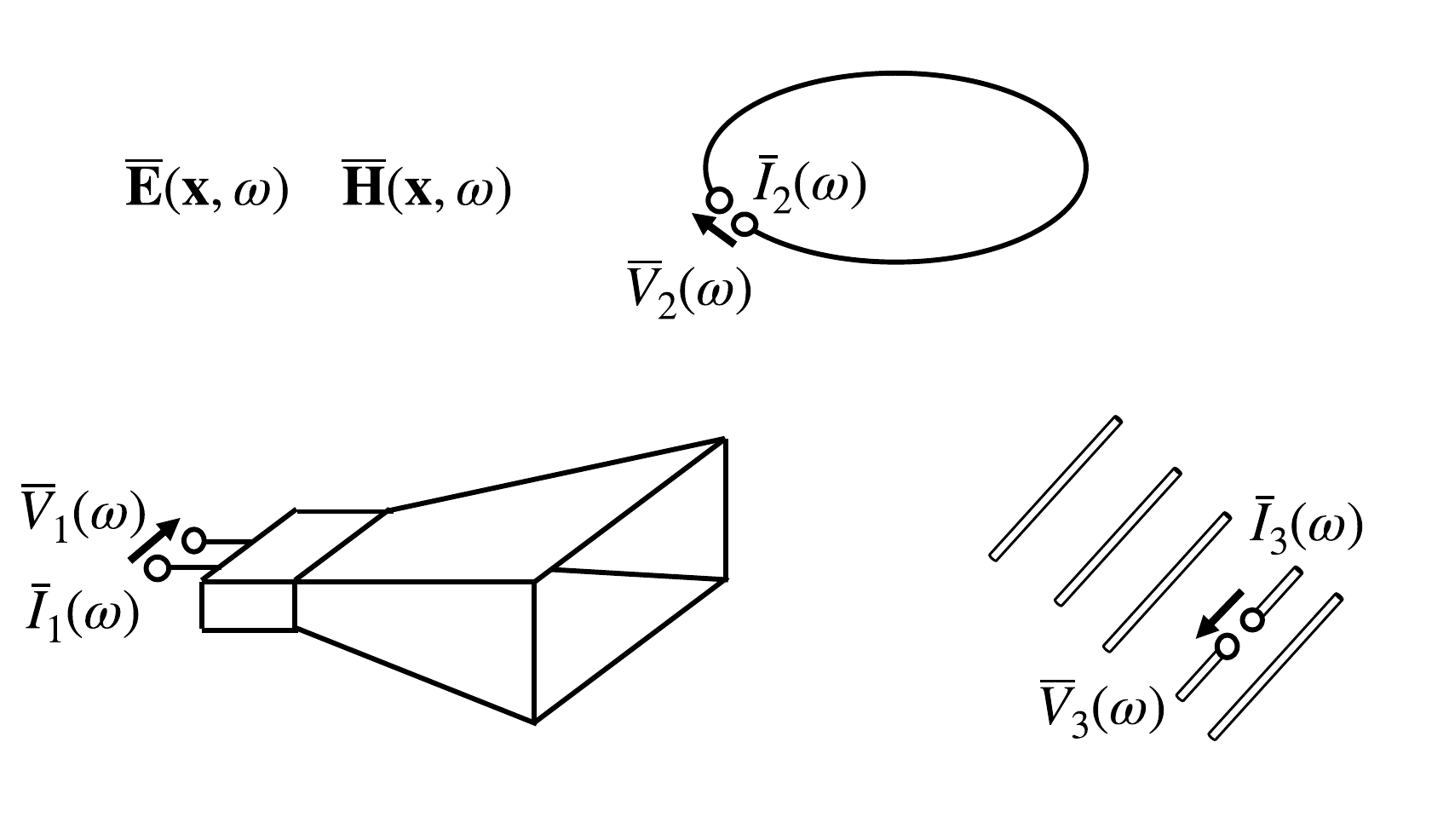}
  \caption{Three different antennas driven by point-like current sources.  a) The current sources $I_n(\om)$ create voltages $V_n(\om)$ b) The current sources $\ov I_n(\om)$ create voltages $\ov V_n(\om)$.   }
  \label{antenna_reciprocity}
  \end{center}
\end{figure}

Fig.~\ref{antenna_reciprocity} shows an arrangement of three different antennas, driven by current sources $I_n(\om)$. These currents result in voltages $V_n(\om)$ across the terminals of the antennas, while the currents $\ov I_n(\om)$ generate a different set of voltages $\ov V_n(\om)$. As before, the Lorentz reciprocity theorem relates these two states and we have
\beq
       \sum_{n=1}^{N=3} I_n(\om) \ov V_n(\om) =  \sum_{n=1}^{N=3} \ov I_n(\om) V_n(\om) 
\eeq
If we again assume just the first antenna to be driven by a current $I_1(\om) = I(\om)$, and in the situation we assume the second antenna to be driven by the same current $\ov I_2(\om)=I(\om)$, we find that $V_2(\om)=\ov V_1(\om)$.
Since this relationship holds for arbitrary antenna geometries and arbitrary relative orientations, we deduce the remarkable result that the reception and transmission characteristics of each antenna must be identical. This is called the antenna reciprocity theorem.

\clearpage 
\newpage

%%%%%%%%%%%%%%%%%%%%%%%%%%%%%%%%%%%%%%%%%%%%%%%%%%%%%%%%%%%
%%%%%%%%%%%%%%%%%%%%%%%%%%%%%%%%%%%%%%%%%%%%%%%%%%%%%%%%%%%

\section{Signals induced by a moving point charge}

\begin{figure}[ht]
 \begin{center}
 a)
  \includegraphics[width=7cm]{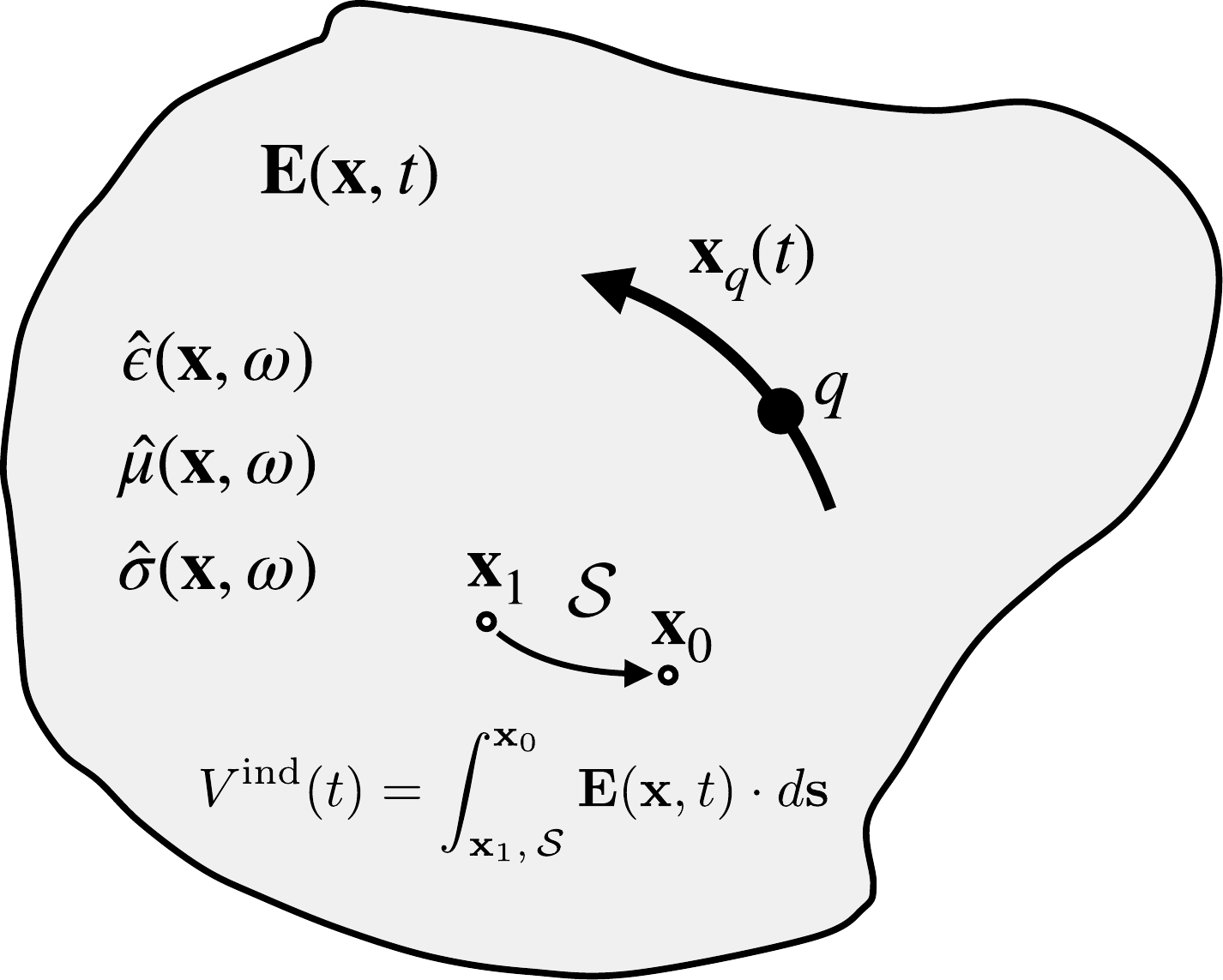}
 b)
    \includegraphics[width=7cm]{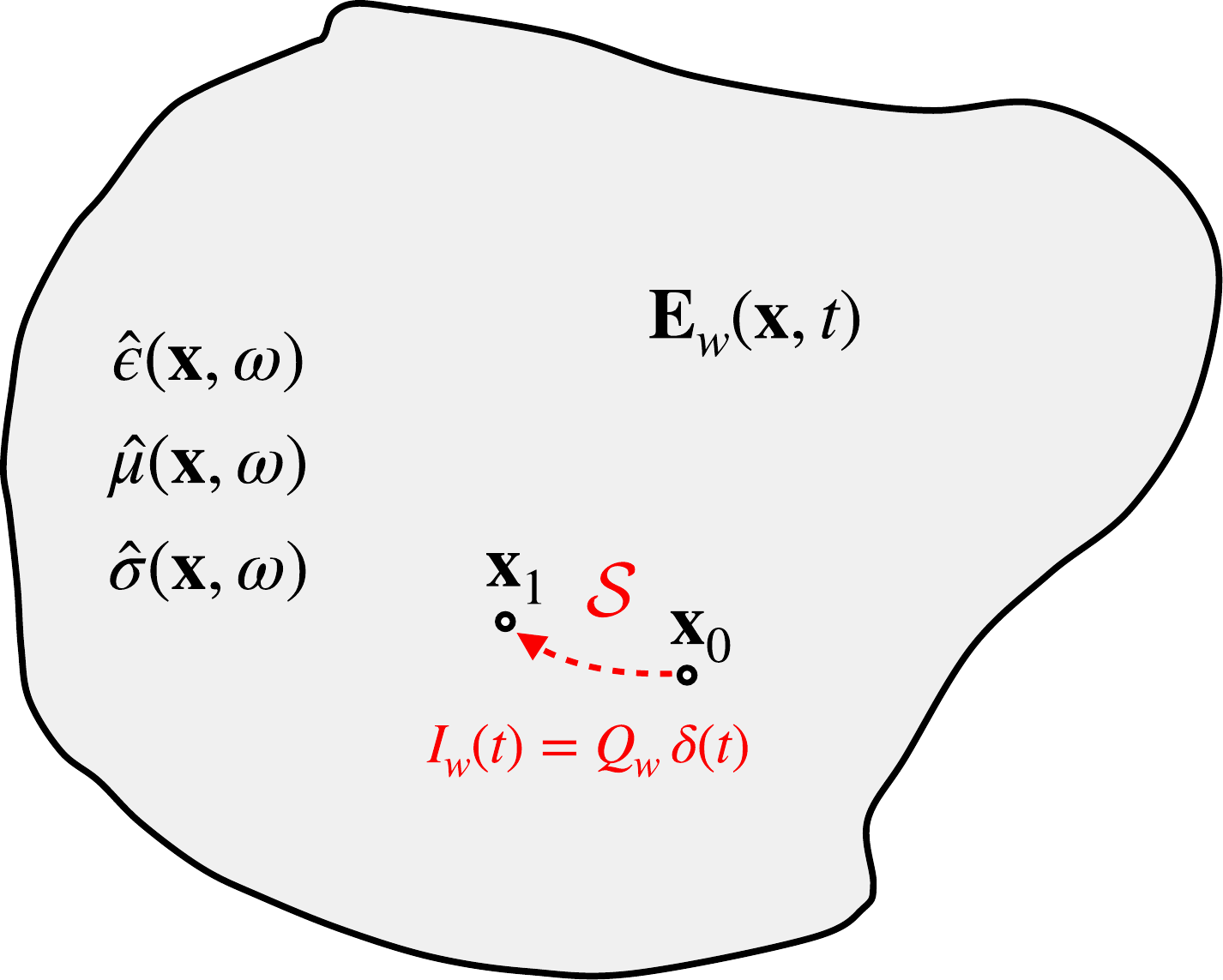}
  \caption{a) A moving point charge is creating an electric field and therefore a 'potential difference' between the points  $\vx_1$ to $\vx_0$. b) A line current $I_w=Q_w \delta(t)$ producing an electric field $\vE_w(\vx, t)$, the so called 'weighting field'.}
  \label{theorem1}
  \end{center}
\end{figure}
We now turn to the signal created by the movement of a point charge $q$ as shown in Fig.~\ref{theorem1}. Electrodes, wires and other detector elements can be absorbed into the material parameters $\hat \vep(\vx), \hat \mu(\vx), \hat \sigma(\vx)$. Let us assume that the point charge $q$ moves along an arbitrary trajectory $\vx_q(t)$, as shown in Fig.~\ref{theorem1}a). The current density created by this motion is given by
\beq
    \vJ^e(\vx, t) = q \dot \vx_q(t) \delta[\vx - \vx_q(t)]
\eeq
We also assume that the detector setup implied by the material distribution delivers its readout signal at the positions $\vx_1$ and $\vx_0$. The integral of the electric field along a particular path $\mathcal{S}$ connecting $\vx_1$ and $\vx_0$ is a quantity related to a potential difference $V^{\mathrm{ind}}(\om )$ between these two points, and therefore to our detector signal. If we parametrize $\mathcal{S}$ as $\vx_s(s)$ with $\vx_s(s_1) = \vx_1$ and $\vx_s(s_0) = \mathbf{x}_0$, the signal can be defined as
\beq
    V^{\mathrm{ind}}(\om) := \int_{\vx_1,\,\mathcal{S}}^{\vx_0}\vE(\vx, \om)\cdot d{\bf s} = \int_{s_1}^{s_0}\vE(\vx_s(s), \om) \frac{d \vx_s(s)}{d s}ds.
\eeq
In general $V(\om)$ is not independent of the path between $\vx_1$ and $\vx_0$ because $\grad \times \vE \neq 0$. In a traditional particle detector the point $\vx_1$ would represent the signal electrode and the point $\vx_0$ would represent the ground reference. 
For the second situation, shown in Fig.~\ref{theorem1}b), we remove the point charge and place a line current $I_w(\om)$ flowing from $\vx_0$ to the detector terminal at $\vx_1$ along the same path $\mathcal{S}$. The direction of the current is chosen for consistency with conventional definitions. This current will create the weighting field $\vE_w$ as solution to Maxwell's equations. Inserting the above expressions into the reciprocity relation we have
\beq \label{theorem_fourier}
    V^{\mathrm{ind}}(\om) =  \int_{\vx_1,\,\mathcal{S}}^{\vx_0} \vE(\vx, \omega)d{\bf s} = -\frac{1}{I_w(\om)} \int_V  \vE_w(\vx, \om) \vJ^e(\vx, \om) dV
\eeq
If we assume the current $I_w$ to be independent of the frequency $\om$, it corresponds to a delta-like current $I_w(t) = Q_w\delta(t)$ in the time domain. We can then perform the inverse Fourier transform of the above expression and are left with
\beq \label{final_formula}
      V^{\mathrm{ind}}(t)  = -\frac{q}{Q_w} \int_{-\infty}^\infty \vE_w(\vx_q(t'), t-t')\dot \vx_q(t') dt'
\eeq
This is our desired theorem which has the same form as the one given in \cite{radeka} and \cite{werner1}, but is now shown to hold for the full extent of Maxwell's equations and is therefore applicable to a very wide range of detector types using signals induced on electrodes by the movement of charges:
\\ \\
{\bf Theorem:}
\\
{\it 
   The potential difference $V^{\mathrm{ind}}(t)$ integrated along a path $\mathcal{S}$ from $\vx_1$ to $\vx_0$ that is induced by the movement of a point charge $q$ along a trajectory $\vx_q(t)$ can be calculated in the following way: the charge $q$ is removed and a delta current pulse $Q_w \delta(t)$ is placed along $\mathcal{S}$ flowing from $\vx_0$ to $\vx_1$. The response to this current is the electric weighting field $\vE_w (\vx, t)$. The detected voltage signal can be calculated by convolving the weighting field with the velocity of the particle according to Eq.~\ref{final_formula}.
}
\\ \\
In case we are not dealing with a single charge $q$ that is moving along a trajectory $x_q(t)$ for $-\infty<t<\infty$, but we are assuming two charges $q, -q$ that are created at a single point at $t=0$ and then move from there along trajectories $\vx_a(t)$ and $\vx_b(t)$, the signal is given by
\beq
           V^{\mathrm{ind}}(t)  = -\frac{q}{Q_w} \int_{0}^t \vE_w(\vx_a(t'), t-t')\dot \vx_a(t') dt' + \frac{q}{Q_w} \int_{0}^t \vE_w(\vx_b(t'), t-t')\dot \vx_b(t') dt' 
\eeq 
The detector signal $V^{\mathrm{ind}}$ is usually processed by an amplifier or in general some linear signal processing device with a given transfer function $F(\om)$, resulting the output signal $V^{\mathrm{out}}(\om) = F(\om)V^{\mathrm{ind}}(\om)$. From Eqs.~\ref{theorem_fourier} and \ref{final_formula} we see that the amplifier output signal can also be calculated by using the current source $Q_w F(\om)$ to define the weighting field $\vK_w(\vx, t)$, which will then directly yield the amplifier output signal through
\beq \label{final_formula2}
      V^{\mathrm{out}}(t)  = -\frac{q}{Q_w} \int_{-\infty}^\infty \vK_w(\vx_q(t'), t-t')\dot \vx_q(t') dt'.
\eeq
This has some practical advantages for numerical simulation of the weighting field, which is of course easier for a well defined finite bandwidth.

%%%%%%%%%%%%%%%%%%%%%%%%%%%%%%%%%%%%%%%%%%%%%%%%%%%%%%%
\newpage
%%%%%%%%%%%%%%%%%%%%%%%%%%%%%%%%%%%%%%%%%%%%%%%%%%%%%%%

 \section{Dipole antennas as particle detectors} 

\begin{figure}[ht]
 \begin{center}
 a)
     \includegraphics[height=5cm]{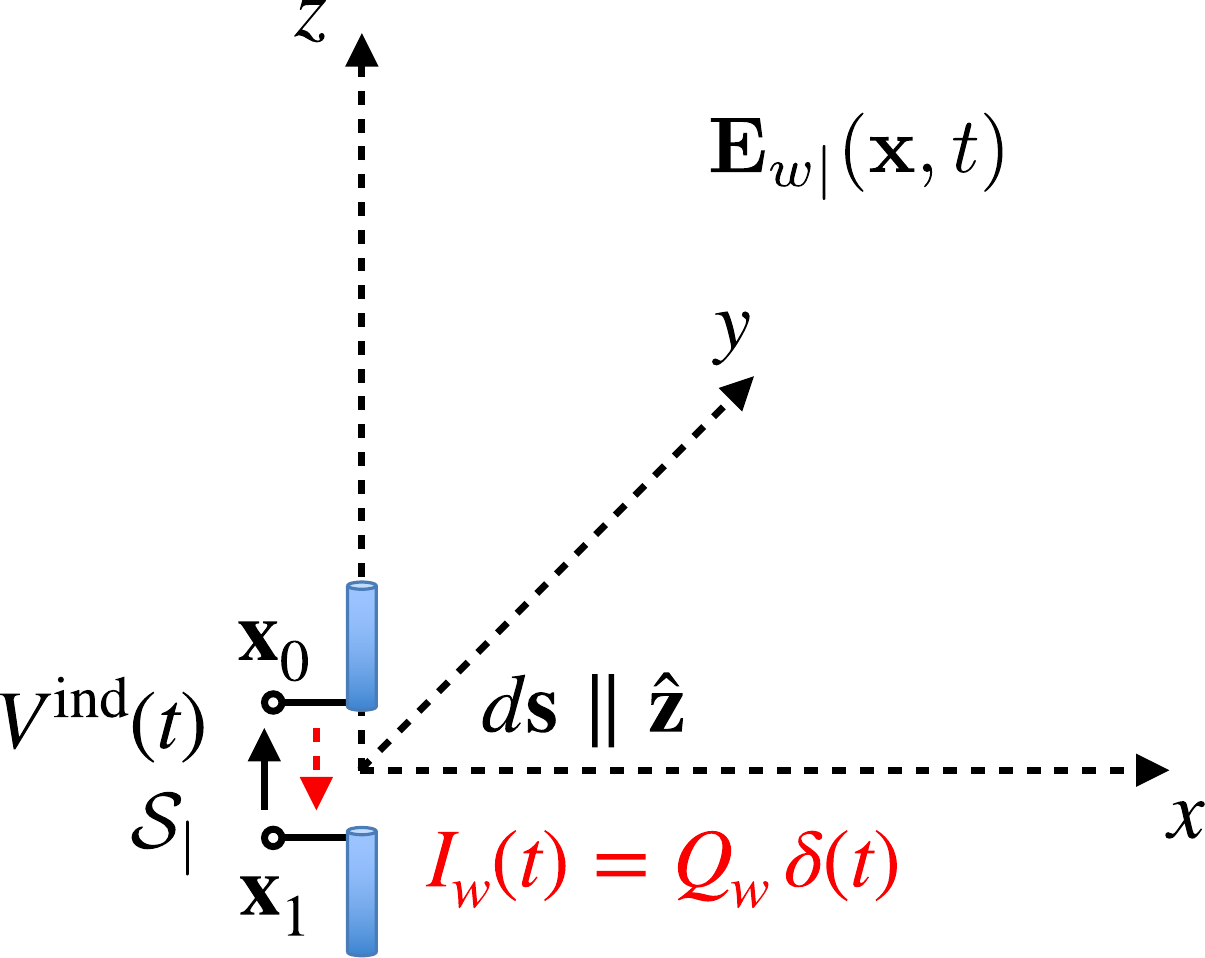}\qquad\qquad
     b)
         \includegraphics[height=5cm]{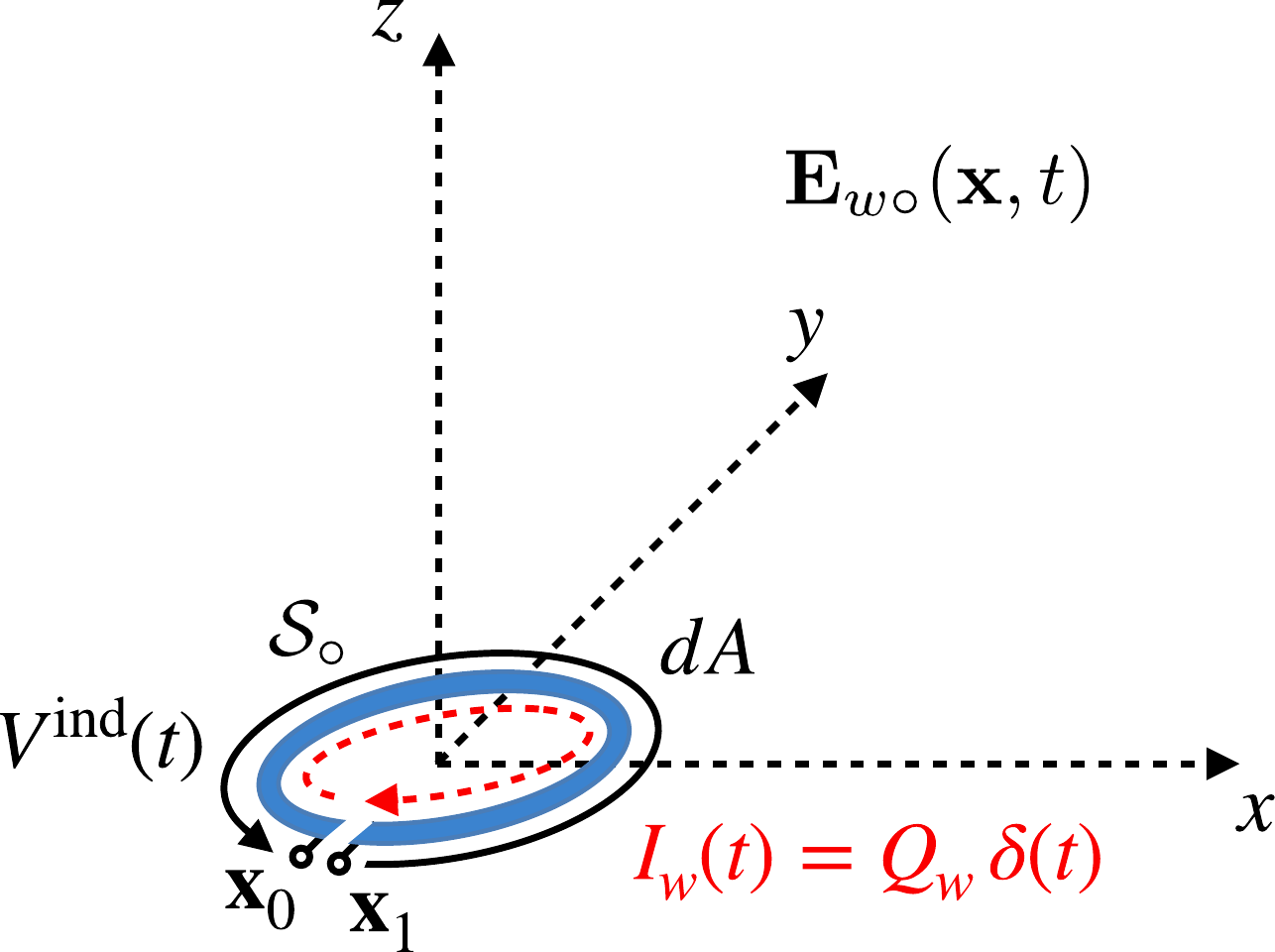}
 \caption{a) The weighting field of an electric dipole antenna. b) The weighting field of a magnetic dipole antenna.}
  \label{dipole_antennas}
  \end{center}
\end{figure}

We now consider two very simple realisations of the general situation outlined in Fig.~\ref{theorem1}. In Fig.~\ref{dipole_antennas}a, we separate the points $\vx_1$ and $\vx_0$ along the $z$-axis by an infinitesimal distance $ds$ and choose as the signal-defining path $\mathcal{S}_{\vert}$ the straight line connecting the two points. We can view $\mathcal{S}_{\vert}$ as connecting two small metal cylinders separated by an air gap of size $ds$. This setup thus forms an infinitesimal electric dipole antenna.
In Fig.~\ref{dipole_antennas}b, we place $\vx_0$ and $\vx_1$ into the $xy$-plane and connect them via an infinitesimal loop $\mathcal{S}_{\circ}$ enclosing an area $dA$. This setup forms an infinitesimal magnetic dipole antenna. In practice the infinitesimal extent means that we are dealing with wavelengths that are much longer than the size of the antennas. \\
To find the weighting field $\vE_{w\vert}$ for the electric dipole antenna, we apply a line current $I_w(t) = Q_w \delta(t)$ along $\mathcal{S}_{\vert}$, as shown in Fig.~\ref{dipole_antennas}a. The weighting field can be computed in the usual way from the electric potential $\varphi$ and the vector potential $\vA$ generated by this current.  Assuming the entire space to be filled with a homogeneous and isotropic medium, its components in spherical coordinates take the form
\bea
    \label{dipole_r}
    E_{w\vert}^\theta(r, \theta) &=& -\frac{Q_w ds}{4\pi \vep} \frac{\sin \theta}{r^3} 
    \left[
        \Theta\left( t - \frac{r \, n}{c}\right) +
       \frac{r \, n}{c} \delta\left( t - \frac{r \, n}{c} \right) +
      \left( \frac{r \, n}{c} \right)^2 \delta' \left( t - \frac{r \, n}{c}   \right)
      \right],\\
    \label{dipole_theta}
    E_{w\vert}^r(r, \theta) &=& -2 \frac{Q_w ds}{4\pi \vep_0} \frac{\cos \theta}{r^3} 
        \left[
        \Theta\left( t - \frac{r \, n}{c}\right) +
       \frac{r \, n}{c} \delta\left( t - \frac{r \, n}{c} \right) 
     \right],\\
    E_{w\vert}^{\phi}(r, \theta) &=& 0,
\eea
where $\Theta(x)$ is the Heaviside step function and $\delta'(x)$ is the distributional derivative of the Dirac delta distribution. The quantity $r$ labels the radial distance from the antenna, $\theta$ is the polar angle measured from the $z$-axis, and $\phi$ is the azimuthal angle measured from the $x$-axis. The propagation speed of the shock front is given by $v = c/n$ where $n =\left(c \, \sqrt{\mu \vep}\right)^{-1}$ is the refractive index of the material and $c$ is the velocity of light in vacuum. Note that for $t > \frac{r \, n}{c}$ the weighting field corresponds to the field generated by a static electric dipole with dipole moment ${\bf p} = -Q_w \,ds\, \hat{\mathbf{z}}$ situated at the origin.

To obtain the weighting field $\vE_{w \circ}$ for the infinitesimal magnetic dipole antenna, we apply a line current $I_w(t) = Q_w \delta(t)$ along the path $\mathcal{S}_{\circ}$ in the direction shown in Fig.~\ref{dipole_antennas}b. A calculation similar to the case of the electric dipole yields
\bea
       E_{w \circ}^{\theta}(r, \theta) &=& 0,\\
       E_{w \circ}^{r}(r, \theta) &=& 0,\\
       E_{w \circ}^{\phi}(r, \theta) &=& \frac{Q_w \, dA \, \mu}{4\pi} \frac{\sin \theta}{r^2}
       \left[
       \delta' \left( t - \frac{r \, n}{c}\right)
      +\frac{r \, n}{c}  \delta'' \left( \ t - \frac{r \, n}{c}\right)
      \right].
\eea
These weighting fields have terms with $1/r^2$-dependence that we typically call 'Coulomb terms' and others with $1/r$-dependence that we call 'radiation terms'. We will use these weighting fields for some concrete examples in the next chapters.

A word is in order about the signal produced by these simple detectors. As expected from the geometry of the path $\mathcal{S_{\vert}}$ that defines the signal, the infinitesimal electric dipole antenna oriented along the $z$-axis simply measures the $z$-component of the electric field produced by a particle moving along an arbitrary trajectory $\vx_q(t)$.
Similarly, the infinitesimal magnetic dipole antenna in the $xy$-plane measures the rate of change of the magnetic flux $dA \cdot B_z$ produced by the moving particle.

For both cases, this can be formally shown by calculating the registered signal through Eq.~\ref{final_formula} starting from the weighting fields $\vE_{w\vert}$ and $\vE_{w\circ}$ and then comparing it to the electric and magnetic fields obtained from the Linard-Wiechert potentials at the place of the antenna. This is shown in Appendix A for the electric dipole and in Appendix B for the magnetic dipole.\\

\begin{figure}[ht]
 \begin{center}
     \includegraphics[width=10cm]{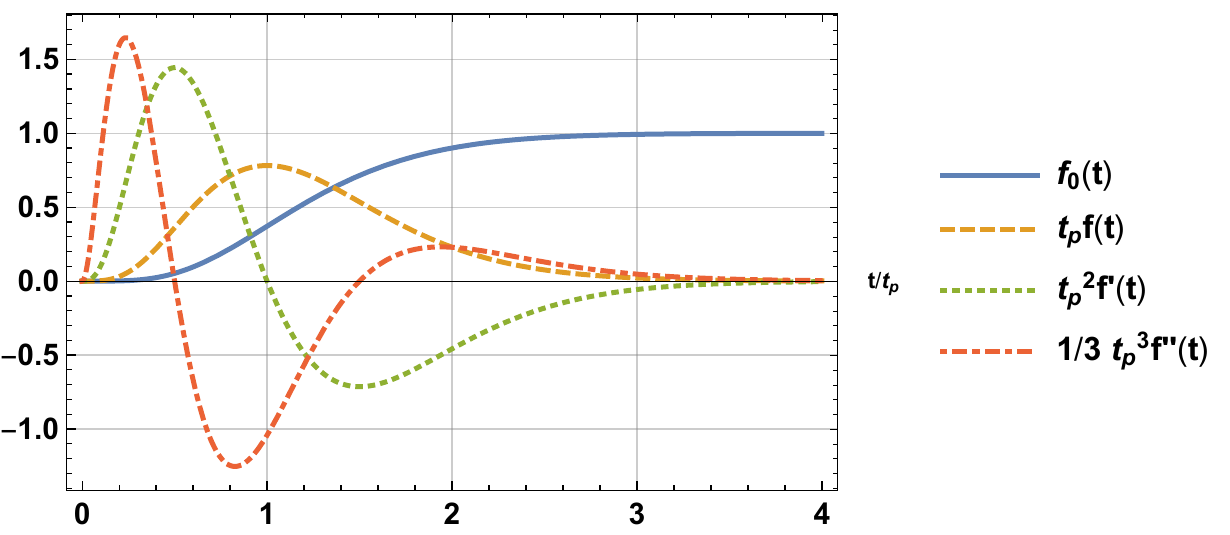}
 \caption{The different terms of the weighting field of an infinitesimal dipole when convoluted with the transfer function of the readout electronics, in this case a low pass filter of order $N=4$.}
  \label{convolution}
  \end{center}
\end{figure}

In case we assume that the detected signal $V^{\mathrm{ind}}(t)$ is further processed by an amplifier with transfer function $F(\om)$ with related delta response $f(t)$, we can directly convolute the weighting field with this transfer function as discussed in the previous section. The non-vanishing components of the resulting weighting fields $\vK_{w\vert}$ for the electric dipole and $\vK_{w\circ}$ for the magnetic dipole then read as
\bea
     K_{w\vert}^{\theta}(r, \theta) & = & -\frac{Q_w \, ds}{4\pi \vep} \frac{\sin \theta}{r^3} 
    \left[
   f_0\left( t - \frac{n \, r}{c}   \right) +  \frac{n \, r}{c} f \left( t - \frac{n \, r}{c}   \right)
      +\left( \frac{n \, r}{c} \right)^2 f' \left( t - \frac{n \, r}{c}   \right)
    \right]  \label{conv1}  \\
    K_{w\vert}^{r}(r, \theta) & = &  -2 \frac{Q_w \, ds}{4\pi \vep} \frac{\cos \theta}{r^3} 
        \left[
        f_0\left( t - \frac{n \, r}{c}   \right) +
      \frac{n \, r}{c} f \left( t - \frac{n \, r}{c}   \right)
    \right]   \label{conv2}  \\
    K_{w\circ}^{\phi}(r, \phi) &  = &  \frac{Q_w \, dA \, \mu}{4\pi} \frac{\sin \theta}{r^2}
       \left[
       f' \left( t - \frac{n \, r}{c}\right)
      +\frac{n \, r}{c} f'' \left( \ t - \frac{n \, r}{c}\right)
       \right] \label{conv3} 
\eea
with $f_0(t)=\int_0^tf(t')dt'$. We use as an example the transfer function of a voltage amplifier with low frequency gain $G$ and peaking time $t_p$ 
\beq
      F(i\om) = \frac{G}{(1+i\om t_p/N)^{N+1}} \qquad    f(t) =  \frac{G}{t_p\,(N-1)! } \left(\frac{Nt}{t_p} \right)^N e^{-N\,t/t_p}   \qquad f_{bw} = \frac{N}{2\pi t_p} \sqrt{2^{1/(N+1)}-1}
\eeq
where $f_{bw}$ is the 3\,dB bandwidth limit of the amplifier and $N=1, 2, 3, ...$ is the order of the filter. Fig.~\ref{convolution} shows the functions $f_0(t), f(t), f'(t), f''(t)$ that appear in Eqs.~\ref{conv1}-\ref{conv3}. We see that these convoluted weighting fields are smooth functions of $r$ and $t$ and therefore very well suited for numerical convolution with whatever particle trajectories.

%%%%%%%%%%%%%%%%%%%%%%%%%%%%%%%%%%%%%%%%%%%%%%%%%%%%%%%%%%%
\newpage
%%%%%%%%%%%%%%%%%%%%%%%%%%%%%%%%%%%%%%%%%%%%%%%%%%%%%%%%%%%

\section{Signals in transmission lines} 

\begin{figure}[ht]
 \begin{center}
 a)
  \includegraphics[height=2.6cm]{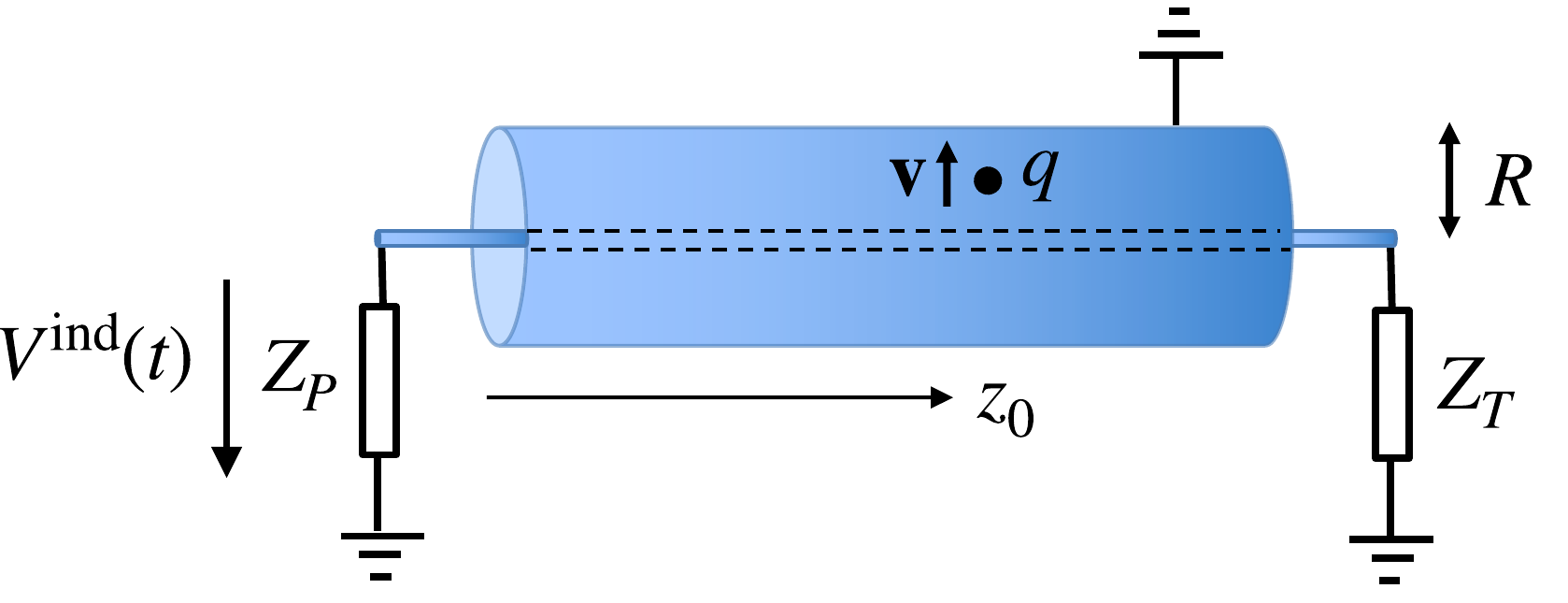}
 b)
    \includegraphics[height=2.6cm]{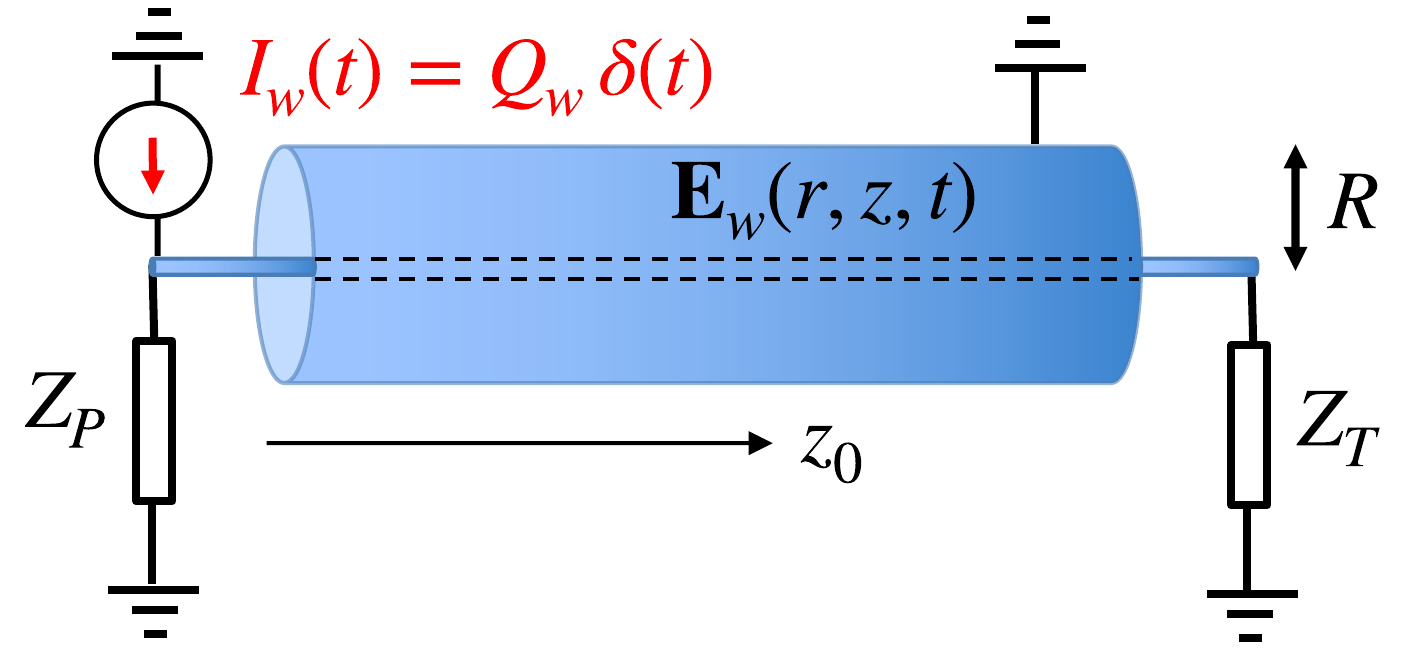}
  \caption{a) A charge $q$ moving inside a transmission line creating a signal $V^{\mathrm{ind}}(t)$ at the input of the readout amplifier. b) The weighting field $E_w(r, z, t)$.}
  \label{transmission_line}
  \end{center}
\end{figure}

Using the formalism developed in the above chapters, we are now in the position to answer the question about signals in transmission lines that was outlined in the introduction. 
Let us assume a coaxial transmission line with wire radius $a$ and tube radius $R$ as shown in Fig.~\ref{transmission_line}a. This setup resembles a wire chamber drift tube. We want to calculate the voltage generated at the amplifier input by a charge moving in radial direction at position $z=z_0$. We assume that only TEM field modes are propagating along the transmission line, which is true for frequencies lower than $c/R$. In this case there are no electric field components along the $z$-direction. The potential difference between the conductors can then be defined in a unique way by $V(z, t) = \int_a^R E(r, z, t) dr$ along any path in the $x{-}y$ plane \cite{clayton}. We therefore consider the situation in Fig.~\ref{transmission_line}b and place a delta current source on the  electrode to determine the weighting field $\vE_w(r, z, t)$ inside the transmission line. The TEM approximation allows this electric field to be written as 
\beq
      \vE_w(r, z, t) = V(z, t) \frac{ \vE^0_w(r)}{V_w} \qquad \vE^0_w(r) = -\grad \varphi (r) \qquad \varphi (r) \vert_{r=a} = V_w \qquad  \varphi (r) \vert_{r=R} = 0
\eeq
The voltage $V(z, t)$ is determined by the transmission line equations for the given stimulus and the potential $\varphi (x, y)$ corresponds to the two dimensional static weighting potential of the central conductor.
Through Eq.~\ref{final_formula}  we have
\beq
        V^{\mathrm{ind}}(t) = -\frac{q}{Q_w} \int_{-\infty}^\infty V(z_0, t-t') \frac{1}{V_w} \vE_w^0 ( r(t') ) \dot \vx(t') dt' =  \frac{1}{Q_w} \int_{-\infty}^{\infty} V(z_0, t-t') I_0(t') dt'
\eeq
The current $I_0(t)$ corresponds to the induced current on the grounded electrode  calculated with the Ramo-Shockley theorem. Writing this expression in the frequency domain gives
\beq
      V^{\mathrm{ind}}(\om) =I_0(\om) \frac{V(z_0, \om)}{Q_w} = I_0(\om) Z(\om) 
\eeq
The impedance $Z(\om)$ relates the voltage $V^{\mathrm{ind}}(\om)$ to the current $I_0(\om)$ at position $z_0$. Using the network reciprocity from Section \ref{network_reciprocity} we have therefore proven the following theorem: \\ \\
{\it  The voltage induced on a transmission line by a charge moving in the $x{-}y$ plane at position $z_0$ can be calculated by first calculating the induced current on the grounded electrode with the electrostatic two dimensional weighting field and then placing this current as an ideal current source one the transmission line at position $z_0$}.  
\\ \\
In case the charge movement also has a $z-$component, it is more practical to first calculate the voltage signal $V(z, t)$ and then to perform the convolution
\beq
        V^{\mathrm{ind}}(t) = \frac{1}{Q_w} \int_{-\infty}^\infty V(z(t'), t-t') I_0(t') dt'.
\eeq

%%%%%%%%%%%%%%%%%%%%%%%%%%%%%%%%%%%%%%%%%%%%%%%%%%%%%%%
\newpage
%%%%%%%%%%%%%%%%%%%%%%%%%%%%%%%%%%%%%%%%%%%%%%%%%%%%%%%

\section{Synchrotron radiation from gyrating electrons} 

\begin{figure}[ht]
 \begin{center}
  \includegraphics[height=5cm]{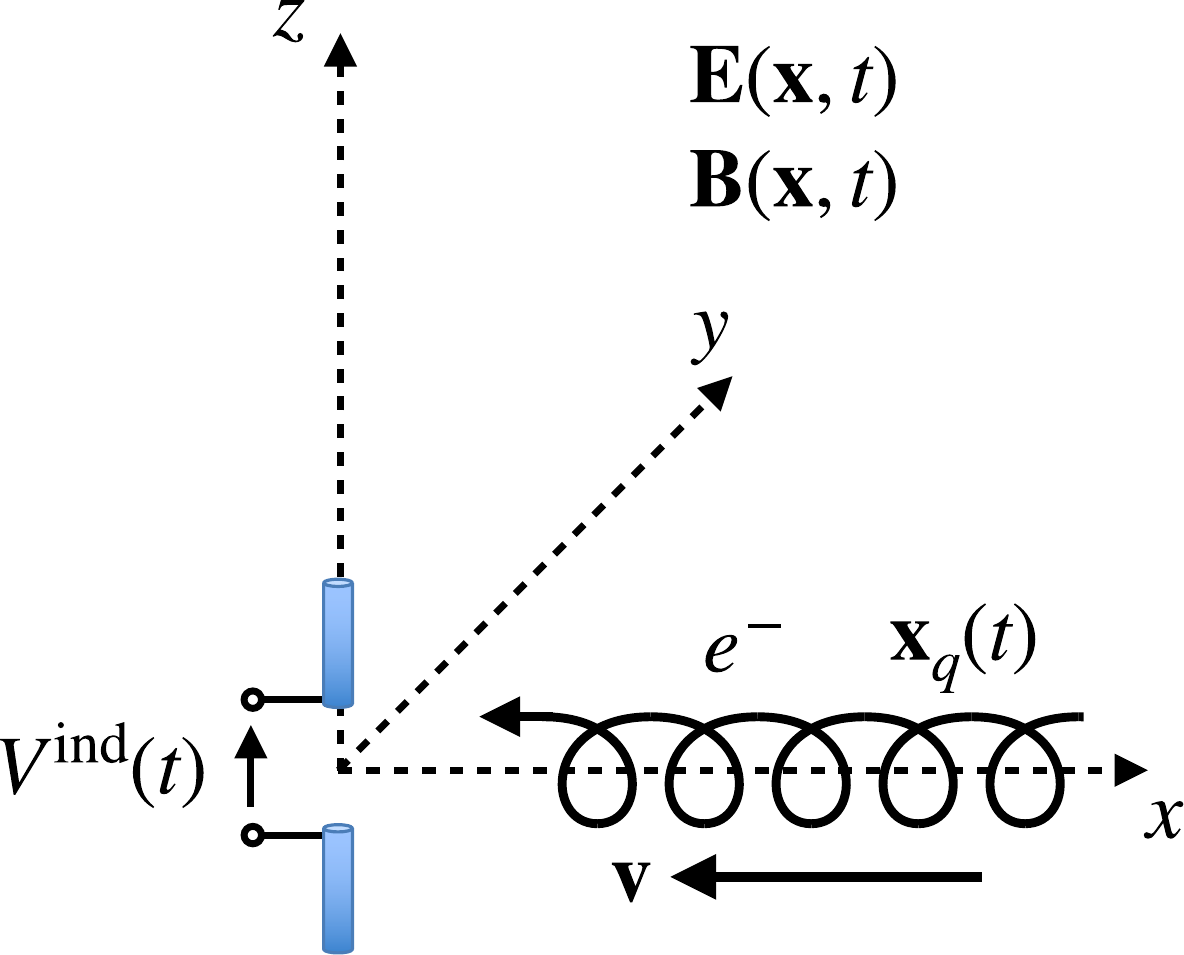}
   \caption{An electron gyrating in a static magnetic field produces radiation that is detected by a dipole antenna. 
  }
  \label{theorem2}
  \end{center}
\end{figure}
As another example for the general applicability of our signal theorem, we use an electric dipole antenna to detect the radiation from an electron that is gyrating in the galactic magnetic field as shown in Fig.~\ref{theorem2}. We assume that the electron moves with velocity $v$ in the negative $x$-direction. Its trajectory and velocity vector are given by
\beq
     \vx_q(t)       = 
      \left(
       \begin{array}{c}
          -vt\\ A \sin (\om_0 t) \\ A \cos (\om_0 t) 
       \end{array}
       \right)
      \qquad  
    \dot \vx_q(t)       = 
      \left(
       \begin{array}{c}
          -v\\ A \om_0 \cos (\om_0 t) \\ - A \om_0 \sin (\om_0 t) 
       \end{array}
       \right)
       \qquad
       r(t) \approx \vert -vt \vert
\eeq
where the frequency $\om_0$ of its gyrating motion is given by the cyclotron frequency $\om_0= e_0 B/m_e$.
Since the distance $r$ to the electron is large, we only keep the $1/r$ term of the weighting field, and along the $x-$direction we have (we are assuming that the antenna and the electron are in vacuum)
\beq
     E_{w\vert}^x=0 \quad E_{w\vert}^y = 0 \qquad E_{w\vert}^z(r, t)  \approx  \frac{Q_w \, ds}{4\pi \vep_0 c^2} \frac{1}{r} 
   \delta' \left( t - \frac{r}{c}   \right).
\eeq
For $t<0$, when the electron is moving towards the antenna, we have 
\bea
     V(t) & = & \frac{e_0}{Q_w} \int_{-\infty}^\infty E_{w\vert}^z(v t', t-t' )A \om_0 \sin (\om_0 t') dt' \\ 
      & \approx &  -\frac{A \, e_0 \, ds}{4\pi \vep_0 c^2} \left[\frac{1}{c} \frac{\om_0}{\beta t^2} 
       \sin (\om t) - \frac{1}{c} \frac{\om_0^2 \cos (\om t) }{\beta t(1-\beta)} \right] \\
     & \approx &  \frac{A \, e_0 \, ds}{4\pi \vep_0 } \, \frac{1}{r(t)} \frac{\om_0^2}{c^2}\frac{ \cos (\om t) }{(1-\beta)} \qquad \mbox{for} \qquad \beta \approx 1
\eea
with $\om = \om_0/(1-\beta)$. 
For galactic magnetic fields on the order of 1\,nT, the frequency $\om_0$ is only 176\,Hz. For electrons with a kinetic energy $E$ the observed frequency is larger by a factor $1/(1-\beta) \approx 2(E/m_ec^2)^2$. For an electron with a kinetic energy of 5\,GeV, the frequency is increased by a factor $2\times 10^6$ so we measure radio waves of around 352\,MHz. \\ 
The same expression describes the synchrotron radiation emitted by an electron beam that is passed through an undulator with a wavelength $\lambda_0$ and where the emitted radiation has a wavelength of $\lambda = \lambda_0(1-\beta)$. 

\section{Signals in a beam current transformer}
\begin{figure}[ht]
 \begin{center}
  \includegraphics[height=5cm]{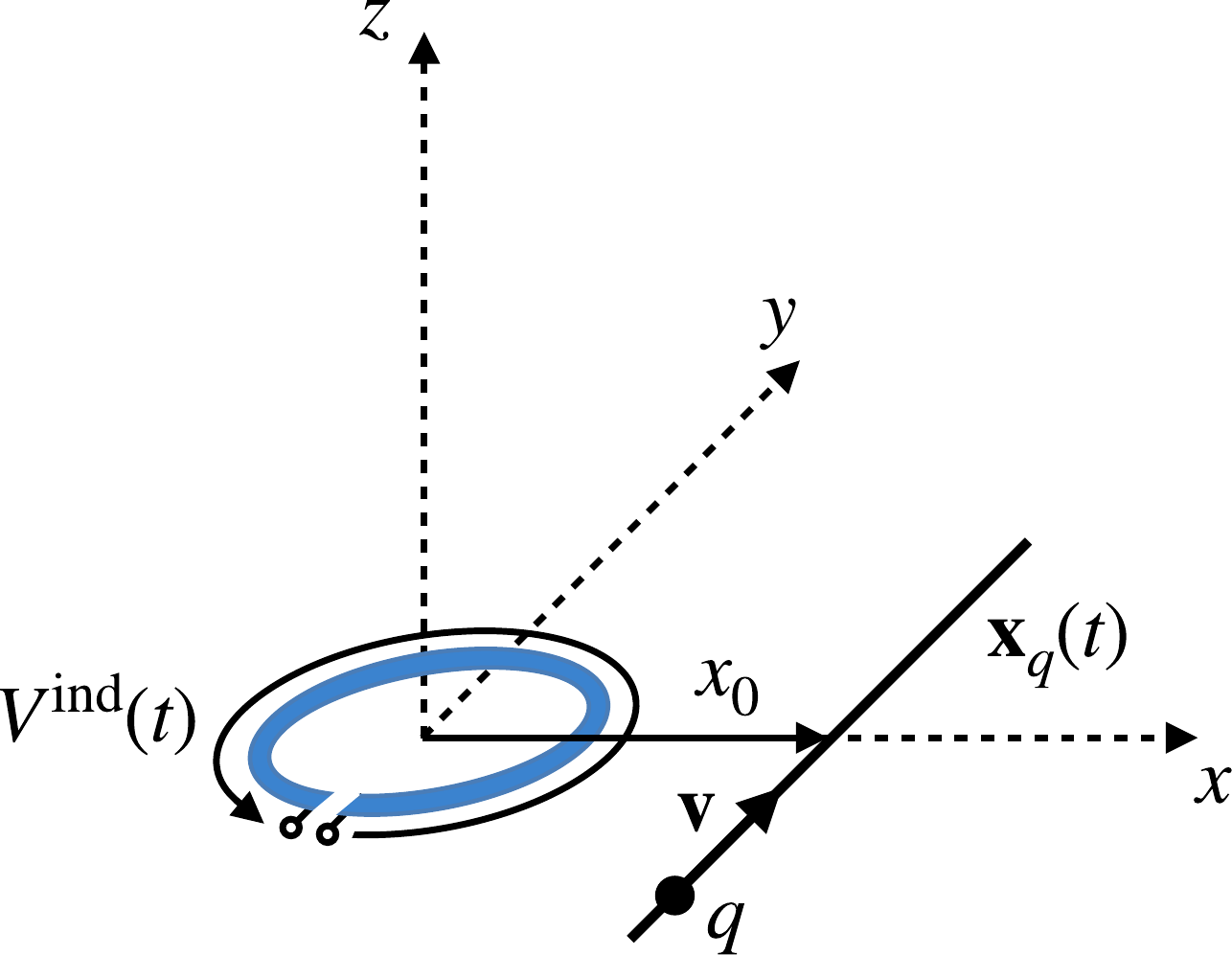}
  \caption{A particle with charge $q$ passing by a current loop (magnetic dipole antenna) inducing a signal $V^{\mathrm{ind}}(t)$. 
  }
  \label{bct}
  \end{center}
\end{figure}

A Beam Current Transformer (BCT) consists of a transformer ring around a particle beam, where the varying magnetic flux in the ring due to the particle beam induces a voltage signal in the windings.  We first assume a small loop situated at the origin and a particle of charge q moving at distance $x_0$ with velocity $v$, as shown in Fig.~\ref{bct}a.

\begin{figure}[ht]
 \begin{center}
  \includegraphics[width=7cm]{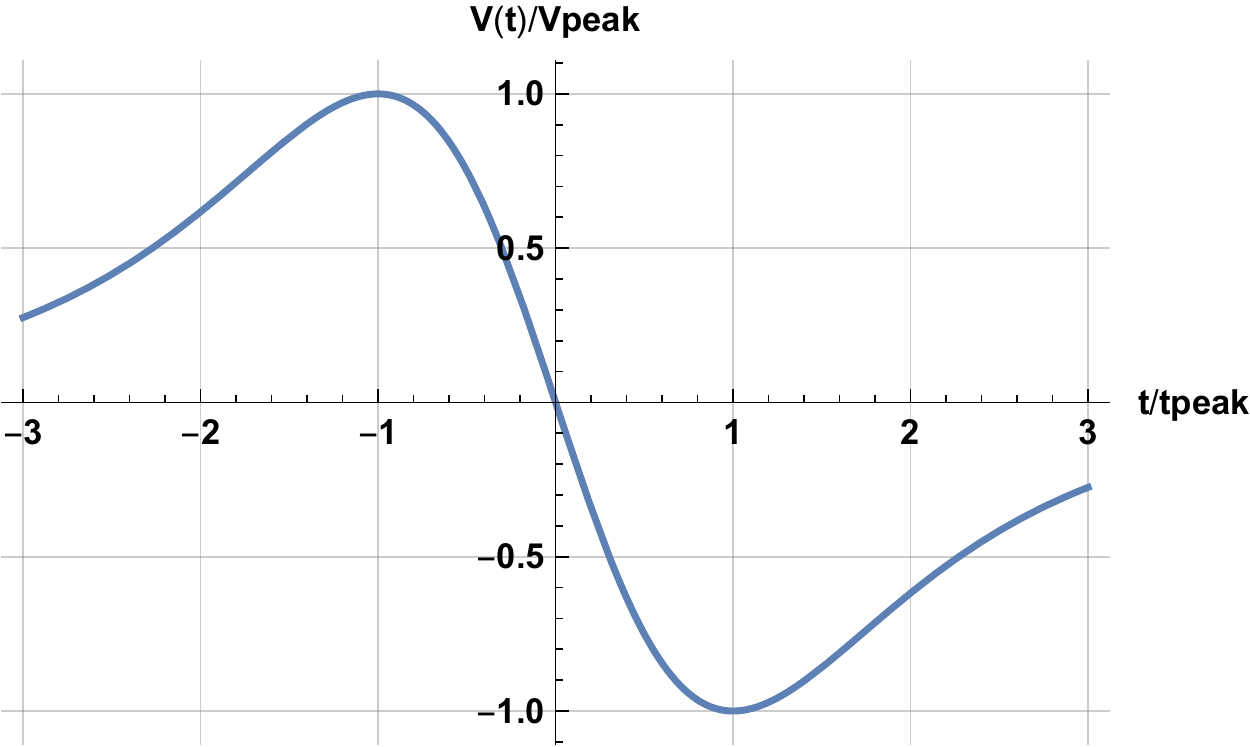}
  \caption{Universal signal shape for a particle passing by a small current loop.}
  \label{bct_signal_shape}
  \end{center}
\end{figure}

We first need the weighting field of a 'small' current loop, which we approximate by an infinitesimal magnetic dipole. Expressed in cartesian coordinates, the $x$ and $y$-components of the weighting field are given by (we again assume $n=1$)
\bea
       E_{w\circ}^{x}(r, \theta, \phi) & = &  -\frac{Q_w \, dA \, \mu_0}{4\pi} \frac{\sin \theta \sin \phi }{r^2}
       \left[
       \delta' \left( t - \frac{r}{c}\right)
      +\frac{r}{c}  \delta'' \left( \ t - \frac{r}{c}\right)
       \right] \label{wf_magdip_x} \\
          E_{w\circ}^{y}(r, \theta, \phi) & = &  \frac{Q_w \, dA \, \mu_0}{4\pi} \frac{\sin \theta \cos \phi}{r^2}
       \left[
       \delta' \left( t - \frac{r}{c}\right)
      +\frac{r}{c}  \delta'' \left( \ t - \frac{r}{c}\right)
       \right] \label{wf_magdip_y}
\eea
The induced voltage $V^{\mathrm{ind}}(t)$ for a movement of the charge along a general trajectory $(x(t), y(t), z(t))$ is 
\beq
    V^{\mathrm{ind}}(t) = \frac{q \, \mu_0 \, dA }{4 \pi } \int \frac{x \dot y-y \dot x}{r^3}
     \left[
         - \delta' \left(   t'-t+\frac{r}{c} \right)  + \frac{r}{c} \delta''  \left(   t'-t+\frac{r}{c} \right)
         \right] dt'
     \label{magnetic_dipole_signal}
\eeq
In our case, the movement of the charge $q$ is parametrised by 
\beq
     x(t) = x_0 \qquad \dot x(t) = 0 \qquad y(t) = vt \qquad \dot y(t) = v \qquad r(t) = \sqrt{x_0^2+v^2t^2}
\eeq
and with some effort we can perform the integration of Eq.~\ref{magnetic_dipole_signal} and get
\bea
     V^{\mathrm{ind}}(t) & = & - \frac{q \, dA}{4 \pi \vep_0} \,\frac{ 3x_0ct \beta^3(1-\beta^2)}{\left[x_0^2(1-\beta^2)+\beta^2c^2t^2\right]^{5/2}}.
\eea
The signal is bipolar and symmetric around the origin. It has two peaks at
\beq
     t_\mathrm{peak} = \pm \frac{x_0}{2 \beta c} \sqrt{1-\beta^2} = \pm \frac{x_0}{2c\beta\gamma} 
\eeq
and the peak value of the signal is
\beq
   V_\mathrm{peak} = \pm \frac{q \, dA}{4 \pi \vep_0} \, \frac{48 \beta^2}{25\sqrt{5}(1-\beta^2) x_0^3}
   = \pm \frac{q \, dA}{4 \pi \vep_0} \, \frac{48 \beta^2\gamma^2}{25\sqrt{5} x_0^3}.
\eeq
For relativistic particles, the peak value of the signal is proportional to $\gamma^2$ of the particle and inversely proportional to the third power of the distance $x_0$. The time of the signal peak $t_\mathrm{peak}$ is proportional to the distance $x_0$ and inversely proportional to $\gamma$ of the particle. The signal, shown in Fig.~\ref{bct_signal_shape}, has a universal shape given by
\beq
    \frac{V^{\mathrm{ind}}(t)}{V_\mathrm{peak}} = \frac{25\sqrt{5}}{32}\, \frac{t/t_\mathrm{peak}}{\left( 1+(t/2t_\mathrm{peak})^2\right)^{5/2}}
\eeq
If we now assume the beam current transformer to be realized by $N$ such loops around the particle beam (i.e.~$N$ windings), the signal $V^{\mathrm{ind}}$ just has to be multiplied by $N$. In case the windings are applied on a ferrite core, the signal must also be multiplied by the permeability $\mu$ of the ferrite. It has to be noted that we did not account for the propagation delay of the signal between the different windings. If we want to take the propagation delay and a possible difference in distance of the particle to the different loops into account, we have to sum the signals with the proper delays instead of just multiplying $V^{\mathrm{ind}}(t)$ by $N$.
\\ \\

%%%%%%%%%%%%%%%%%%%%%%%%%%%%%%%%%%%%%%%%
\newpage
%%%%%%%%%%%%%%%%%%%%%%%%%%%%%%%%%%%%%%%%

\section{Particle shower and Askaryan effect}

\begin{figure}[ht]
 \begin{center}
  \includegraphics[width=8cm]{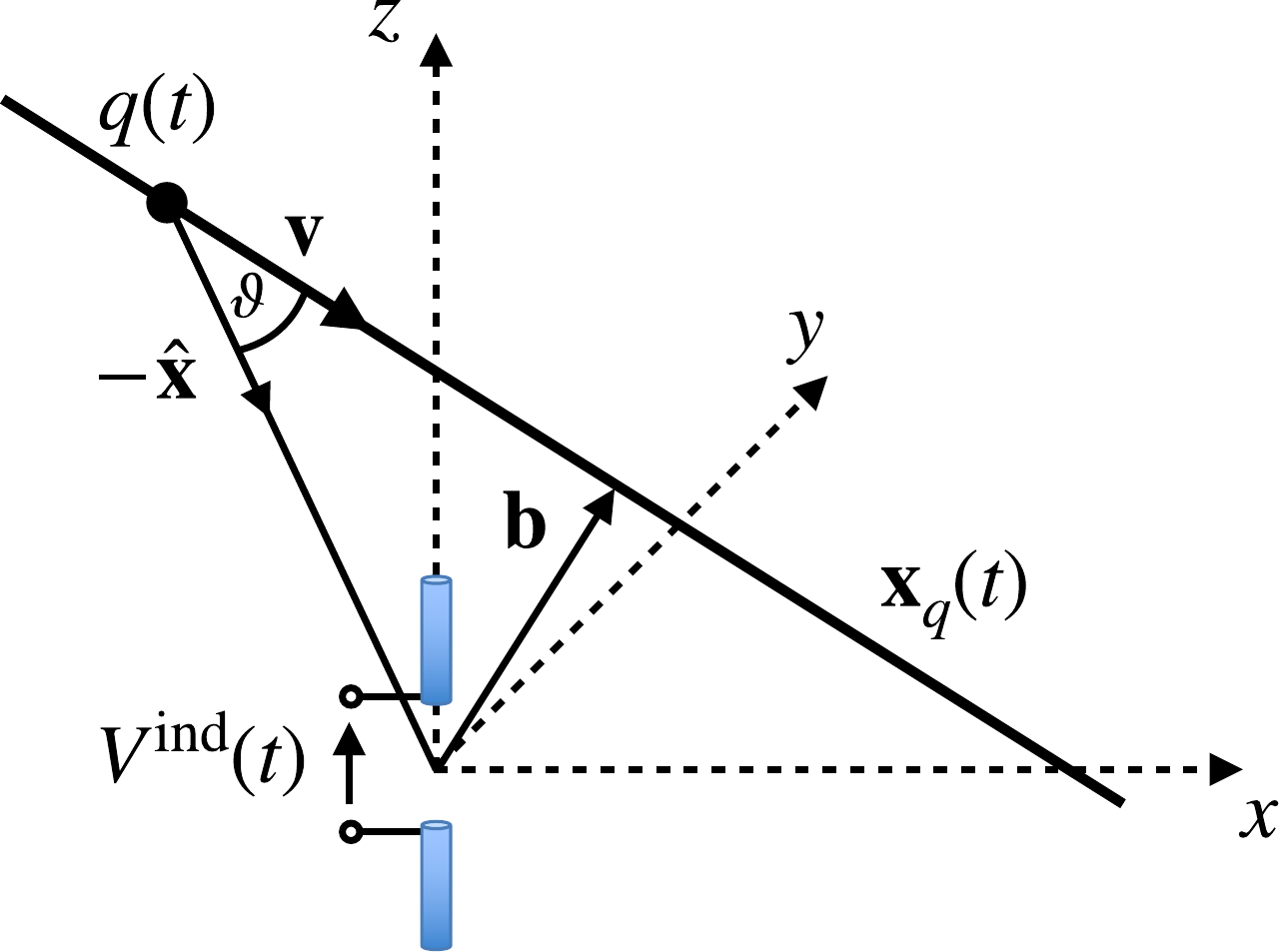}
  \caption{Simple model for an electromagnetic shower. The shower front is represented by a point charge $q(t)$ moving at a constant velocity $\vv$. The radiation produced by the shower is captured by a dipole antenna positioned at the origin.}
  \label{askaryan_shower_setup}
  \end{center}
\end{figure}

The Askaryan effect relates to the fact that in the shower front of an electron-photon shower there is a growing excess of negative charge. This is due to electrons knocked out from atoms in the surrounding medium and due to positron annihilation. Thus, there is an electric current that is the source of electromagnetic radiation, which is exploited in cosmic ray and neutrino experiments. We assume a simple one-dimensional model of a shower, where an exponentially increasing point charge $q(t)$ moves through a medium with refractive index $n$ along a straight trajectory $\vx_q(t) = \vb + \vv t$ with $\vv = \vbeta c$, generating a current $\vJ^{e}(\vx, t) = q(t) \vv \delta(\vx - \vx_q(t))$. We take $q(t) = q \exp\left({\frac{\beta c t}{z_0}}\right)$ and assume $q$ to be negative. The quantity $z_0$ sets the length scale for the development of the cascade. The shower is observed by an infinitesimal electric dipole antenna at the origin, oriented along the $z$-direction. This setup is shown in Fig.~\ref{askaryan_shower_setup}.

Following our signal theorem and performing similar steps as those leading up to Eqs.~\ref{signal_appendix_1}-\ref{signal_appendix_3}, the signal $V^{\mathrm{ind}}$ induced in the dipole antenna is
\bea
V^{\mathrm{ind}}(t) = &-& \frac{ds}{4\pi\ep} \int dt' q(t') \Theta\left(t - t' - \frac{r \, n}{c}\right) \frac{d}{dt'}\frac{z}{r^3}\\
                      &-& \frac{ds}{4\pi\ep} \int dt' q(t') \delta\left(t - t' - \frac{r \, n}{c}\right) \frac{r \, n}{c}\frac{d}{dt'}\frac{z}{r^3}\\
                      &+& \frac{ds}{4\pi\ep} \int dt' q(t') \delta'\left(t - t' - \frac{r \, n}{c}\right) \frac{n^2}{c^2 r^3} \left( z \vx\cdot\dot{\vx} - r^2 \dot{z} \right).
\eea
Assuming that the shower front moves at subluminal velocities, $n \beta < 1$, integrating by parts in the first two terms and performing the integrals leads to
\bea
V^{\mathrm{ind}}(t) = && V^{\mathrm{ind}}_{\mathrm{Front}}(t) + V^{\mathrm{ind}}_{\mathrm{Ions}}(t) + V^{\mathrm{ind}}_{\mathrm{Askaryan}}(t) = \label{total_induced_signal}\\
          = &-& \frac{ds}{4\pi\ep} \left[ \frac{q(t)}{|1 - n \beta \cos\vartheta|^3}\frac{1}{r^2}\left(1 - n^2\beta^2 \right)\left(\frac{z}{r} + n \beta_z \right) \right]_{\tret} -\label{coulomb_contribution}\\
            &+& \frac{ds}{4\pi\ep} \int_{-\infty}^{\tret} dt' \dot{q}(t') \frac{z}{r^3} -\label{static_contribution}\\
            &-& \frac{ds}{4\pi\ep} \left[ \frac{\dot{q}(t)}{|1 - n \beta \cos\vartheta|(1 - n \beta \cos\vartheta)}\frac{n}{r c} \left(n \beta_z + n\beta\cos\vartheta \frac{z}{r} \right) \right]_{\tret},\label{askaryan_contribution}
\eea
where $\vartheta$ is the angle between $\vv$ and $-\hat\vx$ and the retarded time $\tret$ is
\beq
t_{\mathrm{ret}} = \frac{1}{1-n^2 \beta^2} \left[ t - \frac{n}{c} \sqrt{b^2 + \beta^2 (c^2 t^2 - n^2 b^2)} \right].
\eeq
The contribution $V^{\mathrm{ind}}_{\mathrm{Front}}(t)$ in Equation \ref{coulomb_contribution} can be recognised as coming from the (Lorentz-boosted) Coulomb field of the moving charge $q(t)$ representing the shower front, while $V^{\mathrm{ind}}_{\mathrm{Ions}}(t)$ in Equation \ref{static_contribution} represents the (static) Coulomb field produced by the ions created along the trajectory of the shower. The signal component $V^{\mathrm{ind}}_{\mathrm{Askaryan}}(t)$ in Equation \ref{askaryan_contribution} depends on $\dot{q}(t)$ and is commonly referred to as the Askaryan effect \cite{askaryan}.
To explicitly express the induced signal as a function of coordinate time $t$, we can make use of the following relations between retarded and non-retarded quantities,
\bea
\left[ r^2 (1 - n\beta\cos\vartheta)^2 \right]_{\tret} &=& b^2 (1 - n^2 \beta^2) + \beta^2 c^2 t^2, \\
\left[ \cos\vartheta \right]_{\tret} &=& n \beta \left(1 - \frac{c \, t}{\left[ n \, r \right]_{\tret}}\right), \\
\left[ \frac{z}{r} \right]_{\tret} &=& \frac{b_z + \beta_z c t}{\left[ r \right]_{\tret}}- n \beta_z, \\
\left[ r^2 \right]_{t_{\mathrm{ret}}} &=& b^2 + \frac{\beta^2 c^2}{(1-n^2 \beta^2)^2} \left[ t - \frac{n}{c} \sqrt{b^2 + \beta^2 (c^2 t^2 - n^2 b^2)} \right]^2.
\eea

The three terms in Equation \ref{total_induced_signal} can be completed into the following expression for the electric field at the position of the dipole antenna,
\bea
\vE(t) = &-& \frac{ds}{4\pi\ep} \left[ \frac{q(t)}{|1 - n \beta \cos\vartheta|^3}\frac{1}{r^2}\left(1 - n^2\beta^2 \right)\left(\hat\vx + n \vbeta \right) \right]_{\tret} +\\
         &+& \frac{ds}{4\pi\ep} \int_{-\infty}^{\tret} dt' \dot{q}(t') \frac{\vx}{r^3}-\\
         &-& \frac{ds}{4\pi\ep} \left[ \frac{\dot{q}(t)}{|1 - n \beta \cos\vartheta|(1 - n \beta \cos\vartheta)}\frac{n}{r c} \left(n \vbeta - (n \vbeta\cdot\hat\vx) \hat\vx \right) \right]_{\tret} \label{askaryan_contribution_field},
\eea
where it is now seen that the electric field due to the Askaryan effect is oriented along the velocity component $\vbeta_{\perp} = \vbeta - (\vbeta\cdot\hat\vx) \hat\vx$ perpendicular to the line of sight \cite{askaryan_calculation}.

\begin{figure}[ht]
  \begin{center}
    a)
    \includegraphics[height=4.5cm]{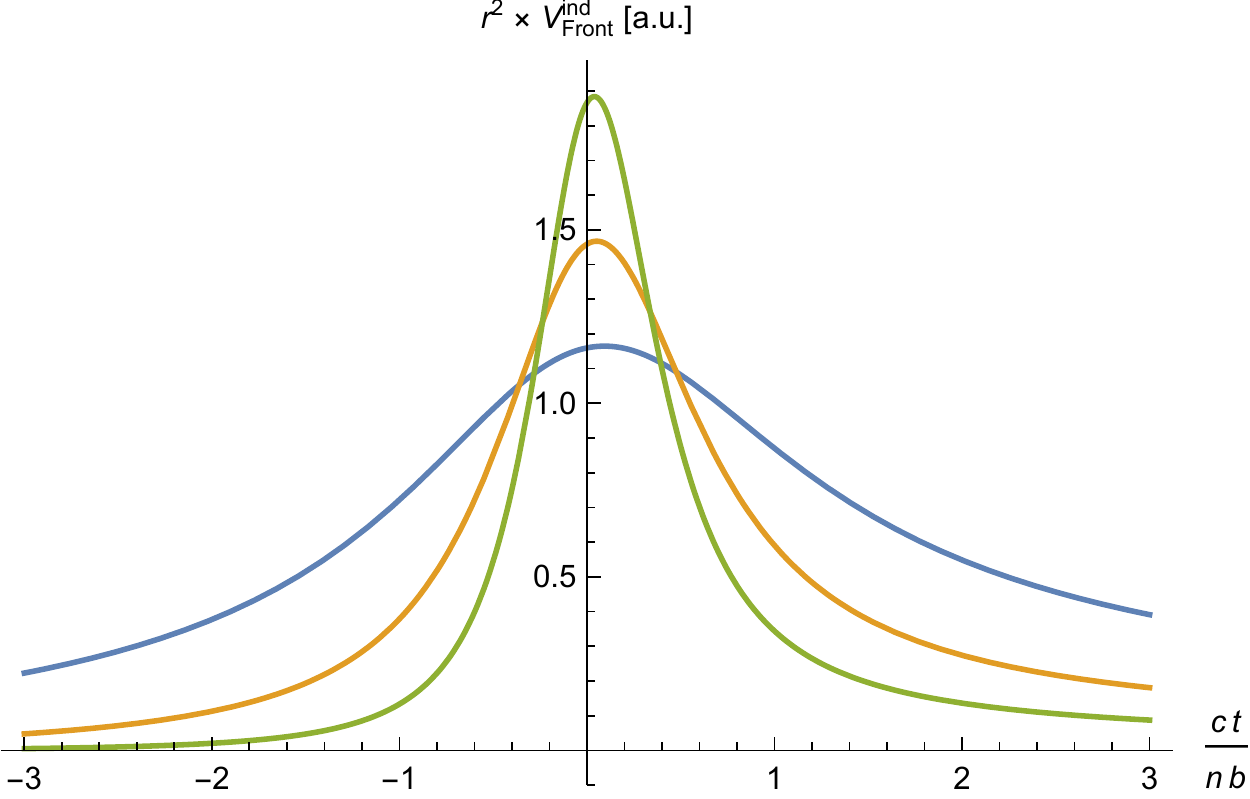}\hspace{0.5cm}
    b)
    \includegraphics[height=4.5cm]{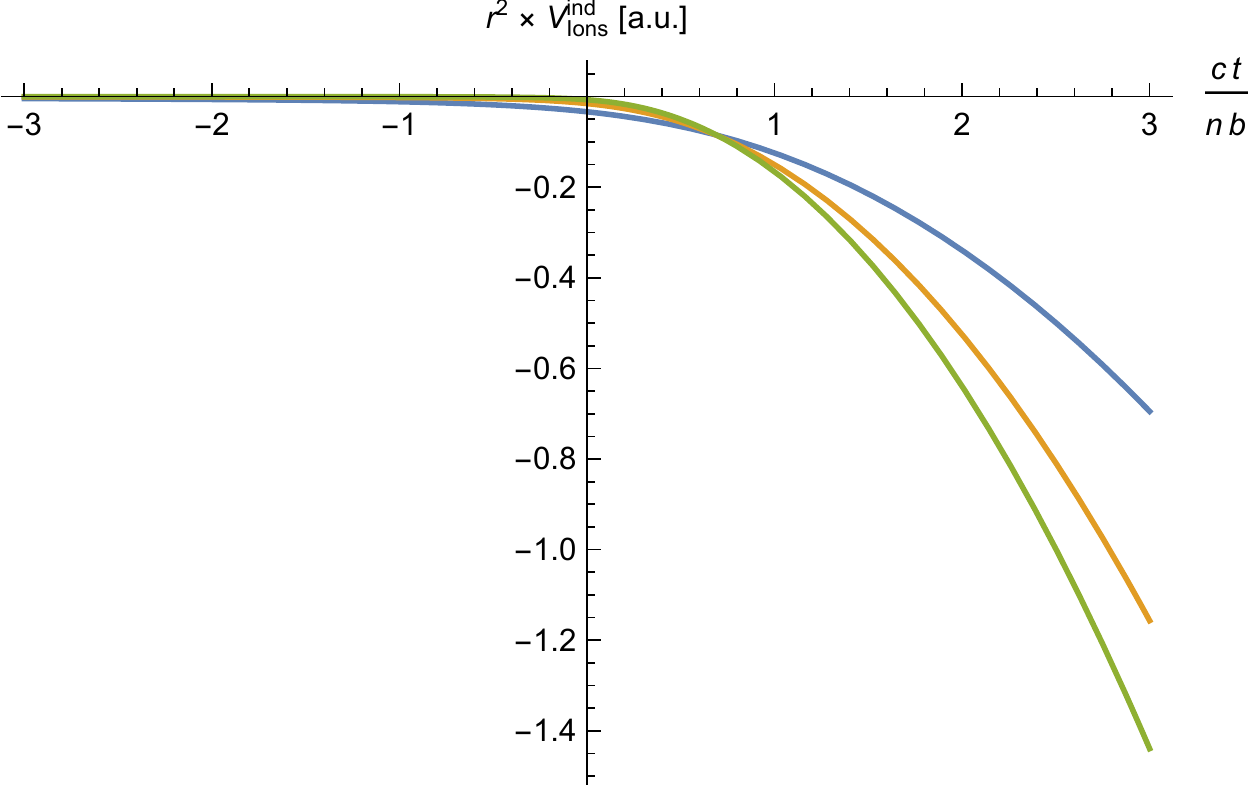}\\\vspace{0.5cm}
    c)
    \includegraphics[height=4.5cm]{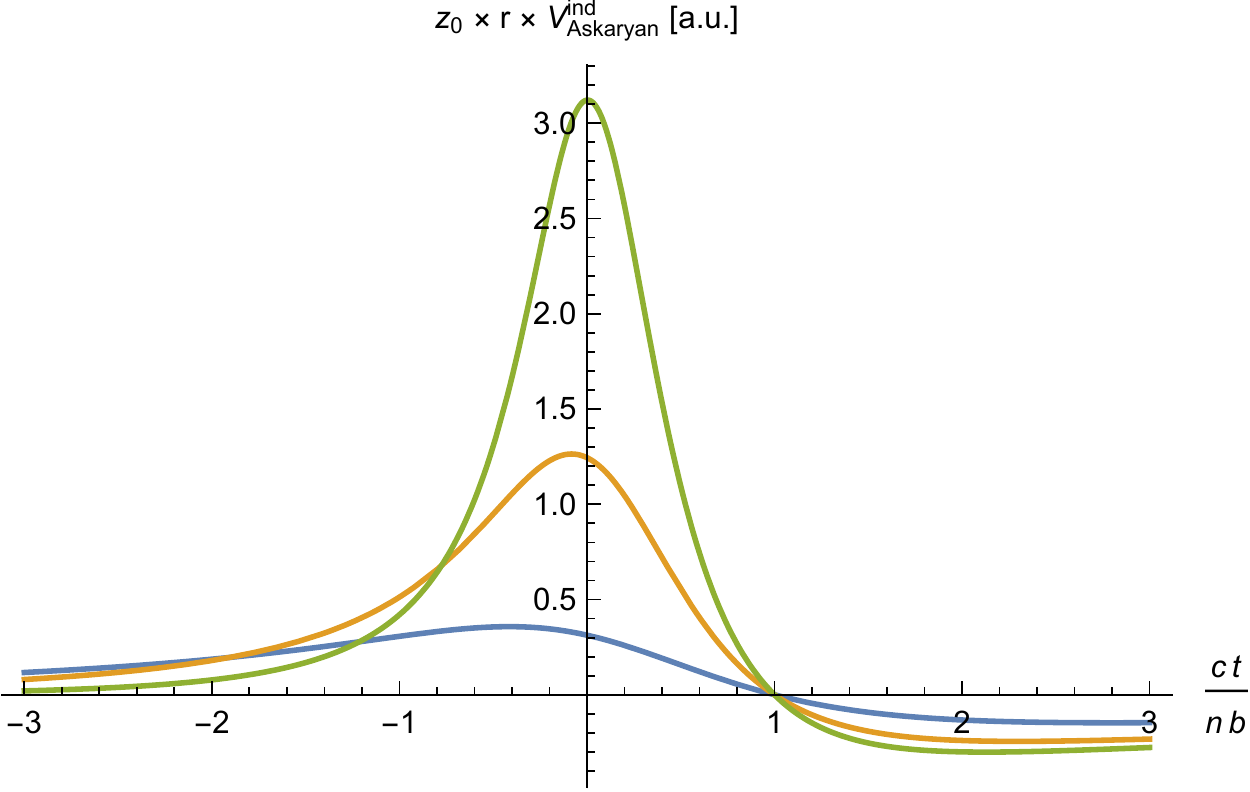}
     \raisebox{1.5cm}{\includegraphics[height=1.5cm]{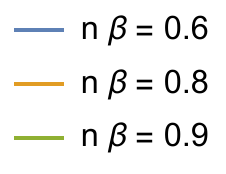}}
  \caption{Signal components induced in the dipole antenna by the developing shower. a) the signal $V^{\mathrm{ind}}_{\mathrm{Front}}$ produced by the Coulomb field of the shower front. b) the signal $V^{\mathrm{ind}}_{\mathrm{Ions}}$ produced by the Coulomb field of the ion tail. c) the signal $V^{\mathrm{ind}}_{\mathrm{Askaryan}}$ attributed to the Askaryan effect. All signal components are suitably scaled by powers of $r(t)$ and $z_0$ to show their intrinsic strength.}
  \label{shower_signal}
  \end{center}
\end{figure}

We visualize the behaviour of the three signal components in Fig.~\ref{shower_signal} for a shower moving parallel to the $x$-axis in the $xz$-plane. To avoid introducing a strong dependency on the late stages of the shower---where our choice for $q(t)$ is not expected to be accurate---we choose $z_0 / b = 10$.
While $V^{\mathrm{ind}}_{\mathrm{Front}}$ and $V^{\mathrm{ind}}_{\mathrm{Ions}}$ lead to unipolar signals, the Askaryan contribution $V^{\mathrm{ind}}_{\mathrm{Askaryan}}$ produces a bipolar signal, arising from a sign flip in $\vbeta_{\perp}$ as the charge moves past the observer. The latter is also seen to become more important at high shower energies.

%%%%%%%%%%%%%%%%%%%%%%%%%%%%%%%%%%%%%%%%
\newpage
%%%%%%%%%%%%%%%%%%%%%%%%%%%%%%%%%%%%%%%%

\section{Conclusion}

We have presented a theorem that allows the calculation of signals generated by moving charges and is valid for the full extent of Maxwell's equations. While the Ramo-Shockley theorem and extensions based on quasi-static approximations are only applicable to traditional particle detectors where the velocity of the charge movement is much smaller that the speed of light, the present theorem is not restricted to these cases. 
It therefore allows the calculation of signals in detectors where signal propagation times and radiation effects are not negligible, like transmission lines and antennas. 
\\ \\
As examples, the signals generated by moving charges in transmission lines and on electric as well as magnetic dipole antennas were presented. The theorem might have important practical applications when performing numerical simulations of signals in particle detectors and on antennas. Instead of explicitly computing the radiation pattern from the charge movement and then calculating the response of a detector to this radiation, one just needs to calculate the weighting field $\vE_w(\vx, t)$ once and for all. One can then derive the detector response to the motion of an arbitrary collection of charges simply by performing the convolution according to Eq.~\ref{final_formula}.

%%%%%%%%%%%%%%%%%%%%%%%%%%%%%%%%%%%%%%%%
%%%%%%%%%%%%%%%%%%%%%%%%%%%%%%%%%%%%%%%%

\section{Acknowledgements} 

We profited greatly from the lecture notes and technical notes by Prof. Kirk T. McDonald, Princeton University.

%%%%%%%%%%%%%%%%%%%%%%%%%%%%%%%%%%%%%%%%
\newpage
%%%%%%%%%%%%%%%%%%%%%%%%%%%%%%%%%%%%%%%%

\section{Appendix A}

In this appendix we show that the electric field due to the arbitrary movement of a point charge $q$ is equal to the signal measured by an electric dipole as calculated through the weighting field formalism. 

\subsection{Electric field for a moving point charge from the Lienard-Wiechert potentials}

We want to find the electric field for a point charge $q$ moving in a medium with refractive index $n$ along trajectory $\vr(t) = (x(t), y(t), z(t))$. The scalar potential $\varphi(\vx, t)$ and vector potential $\vA(\vx, t)$ are given by (Lienard-Wiechert potentials) 
\beq
    \varphi(\vx, t)  = \frac{q}{4 \pi \vep} \int \frac{1}{\vert \vx - \vr(t') \vert} \, \delta \left(   t'-t+\frac{n}{c}\vert  \vx - \vr(t') \vert\right) dt'
\eeq
\beq
  \vA (\vx, t)  = \frac{q \, \mu}{4 \pi } \int \frac{1}{\vert \vx - \vr(t') \vert} \,  \delta \left(   t'-t+\frac{n}{c}\vert  \vx - \vr(t') \vert\right) \dot \vr(t') dt'
\eeq
The electric field is therefore 
\bea
     \vE(\vx, t) & = &   - \grad \varphi(\vx, t) - \frac{\de \vA(\vx, t)}{\de t} = \\
                 & = &   \frac{q}{4 \pi \vep} \int  \frac{\vx-\vr(t')}{\vert \vx - \vr(t') \vert^3}  \delta \left(   t'-t+\frac{n}{c}\vert  \vx - \vr(t') \vert\right) dt' \\
                 & - & \frac{q}{4 \pi \vep} \frac{n}{c} \int  \frac{\vx-\vr(t')}{\vert \vx - \vr(t') \vert^2}  \delta' \left(   t'-t+\frac{n}{c}\vert  \vx - \vr(t') \vert\right) dt' \\
                 & + & \frac{q \, \mu}{4 \pi } \int \frac{1}{\vert \vx - \vr(t') \vert} \,  \delta' \left(   t'-t+\frac{n}{c}\vert  \vx - \vr(t') \vert\right) \dot \vr(t') dt'
\eea
The z-component at the origin is given by 
\bea \label{electric_dipole_field_proof}
   E_z(\vx=0, t) & = &  - \frac{q}{4 \pi \vep} \int  \frac{z(t')}{r(t') ^3}  \delta \left(   t'-t+\frac{n \, r(t')}{c} \right) dt' \\
                 & + & \frac{q}{4 \pi \vep } \frac{n}{c} \int  \frac{z(t')}{r(t')^2}  \delta' \left(   t'-t+\frac{n \, r(t')}{c} \right) dt'  \no \\ 
                 & + & \frac{q}{4 \pi \vep } \frac{n^2}{c^2} \int \frac{ \dot z(t')}{r(t')} \,  \delta' \left(   t'-t+\frac{n \, r(t')}{c} \right)  dt' \no
\eea
with $r(t')$ given by $\vert \vr(t') \vert = \sqrt{x(t)^2+y(t)^2+z(t)^2}$.

%%%%%%%%%%%%%%%%%%%%%%%%%%%

\subsection{Signal on an electric dipole antenna for a moving point charge}

The weighting field for an electric dipole located at the origin and oriented along the z-direction is 
\bea
    E_{w\vert}^x(\vx, t) & = &  \left( E_{w\vert}^r\sin \theta +  E_{w\vert}^{\theta} \cos \theta\right)\cos \phi \\
    \quad 
    E_{w\vert}^y(\vx, t) & = & \left( E_{w\vert}^r\sin \theta +  E_{w\vert}^{\theta} \cos \theta\right)\sin \phi \\
     \quad
    E_{w\vert}^z(\vx, t)  & = &  E_{w\vert}^r\cos \theta -  E_{w\vert}^{\theta} \sin \theta
\eea
where $E_{w\vert}^r, E_{w\vert}^{\theta}$ are from Eq.~\ref{dipole_r} and Eq.~\ref{dipole_theta}. Using 
\beq
    \sin \theta = \frac{z}{r} \quad 
    \cos \theta = \frac{\sqrt{r^2-z^2}}{r} \quad 
    \cos \phi = \frac{x}{\sqrt{r^2-z^2}} \quad 
     \cos \phi = \frac{y}{\sqrt{r^2-z^2}} \label{trig_rels}
\eeq
the signal measured by the dipole antenna is 
\bea
   V^{\mathrm{ind}}(t) & = & -\frac{q}{Q_w} \int  \vE_{w\vert}(\vr(t'),t-t') \dot \vr(t') dt'   \\
        & = & -\frac{q}{Q_w} \int \left[  E_{w\vert}^x(\vr(t'),t-t') \dot x(t')   +  E_{w\vert}^y(\vr(t'),t-t') \dot y(t')   + E_{w\vert}^z(\vr(t'),t-t') \dot z(t')  \right] dt'   \\
        & = & \frac{q \, ds}{4\pi\vep} \int \left(  3 \frac{z\vr \dot \vr}{r^5}-\frac{\dot z}{r^3}  \right)  \Theta \left(  t- t' - \frac{n \, r}{c} \right) dt' \label{signal_appendix_1} \\ 
        & + & \frac{q \, ds}{4\pi\vep} \frac{n}{c} \int \left(  3 \frac{z\vr \dot \vr}{r^4}-\frac{\dot z}{r^2}  \right)  \delta \left(  t- t' - \frac{n \, r}{c} \right) dt' \label{signal_appendix_2} \\ 
        & + & \frac{q \, ds}{4\pi\vep} \frac{n^2}{c^2} \int \left(   \frac{z\vr \dot \vr}{r^3}-\frac{\dot z}{r}  \right)  \delta' \left(  t- t' - \frac{n \, r}{c} \right) dt' \label{signal_appendix_3}
\eea
We use
\beq
   \frac{d}{d t'} \frac{z}{r^3}  =  -3\frac{z\vr \dot \vr}{r^5}+\frac{\dot z}{r^3} 
\eeq        
and rewrite the above expression as
\bea
   V^{\mathrm{ind}}(t)     & = & -\frac{q \, ds}{4\pi\vep} \int \left(   \frac{d}{d t'} \frac{z}{r^3}  \right)  \Theta \left(  t- t' - \frac{n \, r}{c} \right) dt' \\ 
   &- & \frac{q \, ds}{4\pi\vep} \frac{n}{c} \int \left(   r \, \frac{d}{d t'} \frac{z}{r^3}  \right) \delta \left(  t- t' - \frac{n \, r}{c} \right) dt' \\ 
   & + & \frac{q \, ds}{4\pi\vep} \frac{n^2}{c^2} \int \left(   \frac{z\vr \dot \vr}{r^3}-\frac{\dot z}{r}  \right)  \delta' \left(  t- t' - \frac{n \, r}{c} \right) dt'
\eea
Integrating by parts and using the identities
\beq
     \frac{d}{d t'} \Theta \left( t-t'-\frac{n \, r}{c}\right) = -  \delta \left( t-t'-\frac{n \, r}{c}\right)\left( 1+\frac{n}{c}\frac{\vr \dot \vr}{r} \right)
\eeq
and
\beq
    \frac{d}{d t'} \delta \left( t-t'-\frac{n \, r}{c}\right) = -  \delta' \left( t-t'-\frac{n \, r}{c}\right)\left( 1+\frac{n}{c}\frac{\vr \dot \vr}{r} \right)
\eeq
and
\beq
    \delta(x) = \delta(-x) \qquad \delta'(x) = - \delta'(-x) 
\eeq
we have
\beq
 V^{\mathrm{ind}}(t) =  \frac{q \, ds}{4 \pi \vep}
 \int   \left[
            - \frac{z}{r^3}  \delta \left(   t'-t+\frac{n \, r}{c} \right) 
            + \frac{n}{c}\frac{z}{r^2}  \delta' \left(   t'-t+\frac{n \, r}{c} \right) 
            + \frac{n^2}{c^2} \frac{ \dot z}{r} \,  \delta' \left(   t'-t+\frac{n \, r}{c} \right)  
   \right]dt' + \frac{q \, ds}{4 \pi \vep} \frac{z}{r^3}\Big\vert_{-\infty}
\eeq
which, up to a constant term, matches exactly the expression for $E_z(t)$ derived from the Lienard-Wiechert potentials in Eq. \ref{electric_dipole_field_proof}. This offset just reflects the fact that Eq. \ref{final_formula} must evaluate to zero for a static point charge, i.e.~for a case that transcends the validity of the Lorentz reciprocity theorem. Note that this does not represent a limitation of the theorem's practical applicability: any \textsl{static} arrangement of charges can be thought of as being assembled by charges that are slowly brought in from spatial infinity.

%%%%%%%%%%%%%%%%%%%%%%%%%%%%%%%%%%%%%%%%
\newpage
%%%%%%%%%%%%%%%%%%%%%%%%%%%%%%%%%%%%%%%%

\section{Appendix B}

In this appendix we show that the magnetic field due to the arbitrary movement of a point charge $q$ is equal to the signal measured by a magnetic dipole as calculated through the weighting field formalism.

\subsection{Magnetic field for a moving point charge from the Lienard-Wiechert potentials}

The  vector potential $\vA(\vx, t)$ for a point charge $q$ moving in a medium with refractive index $n$ along a trajectory $\vr(t)=(x(t), y(t), z(t))$ are given by (Lienard-Wiechert potentials) 
\beq
  \vA (\vx, t)  = \frac{q \, \mu }{4 \pi } \int \frac{1}{\vert \vx - \vr(t') \vert} \,  \delta \left(   t'-t+\frac{n}{c}\vert  \vx - \vr(t') \vert\right) \dot \vr(t') dt'
\eeq
The magnetic field is therefore 
\bea
     \vB(\vx, t) & = &   \grad\times \vA(\vx, t) \nonumber \\
     & = &  \frac{q \, \mu }{4 \pi } \int \frac{(\vx - \vr) \times \dot \vr}{\vert \vx - \vr \vert^3}
     \left[
         - \delta \left(   t'-t+\frac{n}{c} \vert \vx - \vr \vert \right)  + \frac{n}{c} \vert \vx - \vr \vert \delta' \left(   t'-t+\frac{n}{c} \vert \vx - \vr\vert \right)
     \right] dt' \\\nonumber
\eea
The z-component at the origin is given by
\beq
     B_z(\vx=0, t)
      =   \frac{q \, \mu }{4 \pi } \int \frac{y\dot x - x \dot y}{r^3}
     \left[
       -\delta \left(   t'-t+\frac{n \, r}{c} \right)  + \frac{n \, r}{c} \delta' \left(   t'-t+\frac{n \, r}{c} \right)
     \right] dt' \\\nonumber
\eeq
The time derivative of the magnetic flux through a small area dA around the origin is
\beq \label{magnetic_dipole_proof} 
-dA \frac{dB_z}{dt} = \frac{q \, \mu \, dA}{4 \pi } \int \frac{y \dot x-x \dot y}{r^3}
     \left[
         - \delta' \left(   t'-t+\frac{n \, r}{c} \right)  + \frac{n \, r}{c} \delta''  \left(   t'-t+\frac{n \, r}{c} \right)
     \right] dt' \\\nonumber
\eeq

\subsection{Signal on a magnetic dipole antenna for a moving point charge}%
Starting from the weighting field $\vE_{w\circ}$ in cartesian coordinates given in Eqs.~\ref{wf_magdip_x} and \ref{wf_magdip_y}
we find for the signal
\bea
  V^{\mathrm{ind}}(t) & = & -\frac{q}{Q_w} \int   \vE_{w\circ}\left(\vr(t'),t-t'\right) \dot \vr(t') dt'   \\
        & = & -\frac{q}{Q_w} \int \left[  E_{w\circ}^x(\vr(t'),t-t') \dot x(t')   +  E_{w\circ}^y(\vr(t'),t-t') \dot y(t')  \right]dt'   \\
       & = &  \frac{q \, \mu \, dA }{4 \pi } \int \frac{y \dot x-x \dot y}{r^3}
     \left[
         - \delta' \left(   t'-t+\frac{n \, r}{c} \right)  + \frac{n \, r}{c} \delta''  \left(   t'-t+\frac{n \, r}{c} \right)
     \right] dt' 
\eea
which is identical to the above result for $-dA \frac{dB_z}{dt}$ obtained from the Lienard-Wiechert potentials (Eq. \ref{magnetic_dipole_proof}).

%%%%%%%%%%%%%%%%%%%%%%%%%%%%%%%%%%%%%%%%
\newpage
%%%%%%%%%%%%%%%%%%%%%%%%%%%%%%%%%%%%%%%%

\end{document}